\documentclass[sigconf]{acmart}
\usepackage{graphicx}
\usepackage{amsmath}
\usepackage{booktabs}
\usepackage{algorithm}
\usepackage{algorithmic}
\usepackage{amsfonts}
\usepackage{multirow}
\usepackage{makecell}
\usepackage{subfigure}
\usepackage{color}
\usepackage{bm}
\usepackage{epstopdf}
\usepackage{url}
\usepackage[cal=cm]{mathalfa}
\usepackage{balance}
\usepackage{threeparttable}
\usepackage{lipsum}
\usepackage{enumitem}
\usepackage{pifont}
\usepackage{tabularx}
\usepackage{makecell}
\usepackage{float}
\usepackage[table]{xcolor}
\usepackage{array}
\usepackage{afterpage}

\newcommand{\myred}[1]{\textcolor{red}{#1}}
\newcommand{\myblue}[1]{\textcolor{blue}{#1}}      % 蓝色
\newcommand{\myorange}[1]{\textcolor{orange}{#1}}  % 橙色

\setlist[itemize]{leftmargin=*}

\AtBeginDocument{%
  \providecommand\BibTeX{{%
    \normalfont B\kern-0.5em{\scshape i\kern-0.25em b}\kern-0.8em\TeX}}}

\renewcommand\footnotetextcopyrightpermission[1]{}

\author{Kun Zhang*, Jingming Zhang*, Wei Cheng*, Yansong Cheng*, Jiaqi Zhang*, Hao Lu, Xu Zhang, Haixiang Gan, Jiangxia Cao$^\star$, Tenglong Wang, Ximing Zhang, Boyang Xia, Kuo Cai, Shiyao Wang, Hongjian Dou, Jinkai Yu, Mingxing Wen, Qiang Luo, Dongxu Liang, Chenyi Lei, Jun Wang, Runan Liu, Zhaojie Liu, Ruiming Tang, Tingting Gao, Shaoguo Liu, Yuqing Ding, Hui Kong, \\Han Li, Guorui Zhou, Wenwu Ou, Kun Gai}
\affiliation{
  \institution{Kuaishou Technology, Beijing, China}
 \country{\{zhangkun10, zhangjingming, chengwei07, chengyansong, zhangjiaqi15, luhao, zhangxu20, ganhaixiang03, caojiangxia, wangtenglong, zhangximing, xiaboyang, caikuo, wangshiyao08, douhongjian, yujinkai, wenmingxing, luoqiang, liangdongxu, yangyifan12, wangjun03, liurunnan zhaotianxing, tangruiming, lisize, liushaoguo, dingyuqin03, konghui, lihan08, zhouguorui, luocheng10\}@kuaishou.com, kun.gai@qq.com}
}
\thanks{* Equal Contributions, Jiangxia Cao is the corresponding author.}

\begin{document}
\title{OneMall: One Architecture, More Scenarios — End-to-End Generative Recommender Family at Kuaishou E-Commerce}

\renewcommand{\shorttitle}{OneMall}

\begin{abstract}
In the wave of generative recommendation, we present \textbf{OneMall}, an end-to-end generative recommendation framework tailored for e-commerce services at Kuaishou.
Our OneMall systematically unifies the e-commerce's multiple item distribution scenarios, such as Product-card, short-video and live-streaming.
Specifically, it comprises three key components, aligning the entire model training pipeline to the LLM's pre-training/post-training: (1) \textit{E-commerce Semantic Tokenizer}: we provide a tokenizer solution that captures both real-world semantics and business-specific item relations across different scenarios; (2) \textit{Transformer-based Architecture}: we largely utilize Transformer as our model backbone, e.g., employing Query-Former for long sequence compression, Cross-Attention for multi-behavior sequence fusion, and Sparse MoE for scalable auto-regressive generation; (3) \textit{Reinforcement Learning Pipeline}: we further connect retrieval and ranking models via RL, enabling the ranking model to serve as a reward signal for end-to-end policy retrieval model optimization.
Extensive experiments demonstrate that OneMall achieves consistent improvements across all e-commerce scenarios: +14.7\% GMV in Product-Card, +10.3\% GMV in Short-Video, and +4.9\% GMV in Live-Streaming. OneMall has been deployed, serving over 400 million daily active users at Kuaishou.
\end{abstract}

\maketitle

\section{Introduction}
\myorange{\textbf{LLM Success Lesson}}. As is well known, OpenAI’s scaling laws~\cite{achiam2023gpt} have significantly contributed to technique routing for LLM scaling in natural language processing (NLP).
These laws have directly motivated numerous researchers to pursue the ceiling performance of LLM intelligence, from the data volume, model size, computation resource and reasoning perspectives.
Retrospectively, the LLM success can be attributed to the following progresses:
\begin{itemize}
    \item Pre-training Next-Token Prediction with Large Data-Volume/Model-Parameter: Thanks to the NTP loss unifying all tasks in the same prediction paradigm, therefore LLM could support extensive Trillion-tokens internet training corpus data and model parameter scalings. E.g., evolving from early dense-activation models with 0.03B parameters (such as GPT-2~\cite{radford2019language}) to sparse-activation models containing 671B parameters (like Deepseek V3~\cite{liu2024deepseek}).
    \item Post-training Reasoning Reinforcement Learning: Based on the pre-trained base model, the Reinforcement learning~\cite{schulman2017proximal} further aligns the model's behavior more closely to human taste. E.g., from the early high-quality data SFT to the DPO/GRPO/GSPO series fine-tuning~\cite{rafailov2023direct, guo2025deepseek, zheng2025group}. 
    \item Infrastructure Parallel Techniques: Due to Transformer's elegant GPU-friendly architecture, which has inspired the development of various parallel techniques in distributed systems, including: 
    (1) Fully Sharded Data Parallel (ZeRO series)~\cite{zhao2023pytorch,rajbhandari2020zero}; (2) Tensor Parallel~\cite{shoeybi2019megatron}; (3) Sequence Parallel~\cite{jacobs2023deepspeed}; (4) Pipeline Parallel~\cite{huang2019gpipe}; (5) Context Parallel and so on~\cite{korthikanti2023reducing}, to accelerate the model training/inference efficiency~\cite{kwon2023efficient, zheng2024sglang}.
\end{itemize}
With advances in pre-training, post-training, and infrastructure, the bottleneck of training on tons of Tokens at hundreds of Billion parameters within acceptable time and computation resource limitations has been significantly reduced and possible, enabling the LLM community to continuously incubate advanced artificial intelligence toward human-level performance, rise to many superintelligent systems such as Chat-Bot~\cite{team2025kimi}, Vision-Generator~\cite{wu2025qwen}, Code-Agent~\cite{5team2025glm45agenticreasoningcoding} and others.
Given such success of LLMs, many industrial researchers have been pondering one thing: can we replicate the LLM evolution trace at pre-/post-training in RecSys?

\myorange{\textbf{Background of RecSys}}. In fact, this is a challenging open problem, since the industry RecSys and LLM are not entirely consistent in the task definitions, making it difficult to directly upgrade the model structure~\cite{youtubednn,ma2018modeling} to entire Transformer blocks.
\myblue{For a better understanding the difference, we first introduce the two important recommendation models designing insights in (current) RecSys chain, and their goals}:
\begin{itemize}
    \item Stage 1, Generative retrieval model~\cite{yi2019sampling, cen2020controllable,liu2024crm}: this series of models focuses on utilizing user-side historical interaction logs to predict other daily-hot items that users have not interacted with yet. For instance, retrieve 1,000 items from the total  daily-hot item pool based on user's latest 100 interaction log. In general, it can be seen as multi-targets `Next-Item prediction'.
    \item Stage 2, Discriminate ranking model~\cite{guo2017deepfm, cheng2025choruscvr}: this series of models focuses on estimating the precise probabilities of behaviors that will happen, according to user-side and candidate item-side features. For instance, prediction the click rate (CTR), conversion rate (CVR) and so on, it can be seen as a multiple-objective paradigm to judge which item will attract user more~\cite{cao2025pantheon}.
\end{itemize}
Under these two paradigms with completely different objectives, these two types of models are usually optimized independently of each other, and supporting our recommendation chain in a cascaded manner for a long time, and widely accepted by large companies.
Therefore, the common wisdom in RecSys community has two different scaling technique routes.

\myorange{\textbf{Ranking Model Scaling?}} Roughly speaking, over the past several years, \myred{many well-known advancements in industrial RecSys have come from the Stage 2 discriminative ranking models}, such as how to search the evidence user sub-history log sequence according to item candidate information (e.g., DIN/SIM/TWIN~\cite{zhou2018deep,pi2020search,chang2023twin}) and multiple objective optimization (e.g., MMoE~\cite{ma2018modeling}, HoME~\cite{home}).
We acknowledge that discriminative ranking model served as the core growth engine in past years~\cite{chai2025longer, guan2025make}, especially the cascading GSU/ESU architecture to extend the historical sequence to life-long level.
However, while bringing significant benefits, the current discriminative technique route has its internal limitations:
\begin{itemize}
\item \textbf{Un-balanced computation}: 
Unlike Transformers, where all input tokens have the same parameter space and computation cost~\cite{dao2022flashattention}, e.g, all parameter will interact with all input tokens.
In ranking models, the parameters of different components are always separate from each other.
This leads to a problem where different modules contribute unevenly to the overall computational load. For example, approximately 30\% of the total computation is allocated to TWIN-like ESU modules that require end-to-end lifelong-sequence inference, while only about 10\% is allocated to other ESU modules on top-k sequences inference.
\item \textbf{Complex computation}: unlike the pure generative Transformer-based models with plain computational units that can be split into parallel computing units with balanced loads across distribution training nodes, the discriminative series RecSys model computations workflow is too complex, making it difficult to leverage state-of-the-art parallel computing techniques~\cite{harlap2018pipedream,jiang2024megascale}.
\end{itemize}
With the two limitations, some pioneering works were proposed to scale ranking model, such as MARM~\cite{lv2024marm}, Rank-Mixer~\cite{zhu2025rankmixer}, etc.
However, their architectures are significantly different from the classic Transformer, making it hard to leverage the successful experience from the LLM area directly.

\myorange{\textbf{Retrieval Model Scaling?}} In contrast, the retrieval model's generative `Next-item prediction' paradigm closely aligns with the `next-token prediction' paradigm in LLMs~\cite{sasrec}. 
\myred{The difference is that: the retrieval model always has multiple next ground-truth items, while NLP almost have only one ground-truth token for the given item-sequence/token-sequence.
This is because industrial systems always send a batch of items to user device at same time (e.g., 10 short-videos per request), thus those next-items will share same pre-filling user interaction item sequence in retrieval model.}
In this characteristic, for quite a long time, the retrieval models' research focus has been on modeling the diversity~\cite{li2025multi}.
With the recent advances in Transformer-based sequence modeling, there revealed a promising alternative: diversity can be inherently achieved through controlled exploration (e.g., via adequate temperature) within the generative beam search process~\cite{rajput2023recommender}.
This paradigm shift has motivated researchers to explore auto-regressive generative frameworks for a generative retrieval, which can naturally balance depth (relevance) and breadth (diversity) of interest exploration. 
Currently, two dominant frameworks have emerged:
\begin{itemize}
    \item Multiple Interests ANN-based framework: This is the most common and classical architecture, where user interests are compressed into \myred{multiple embeddings via a user tower}, and item properties are compressed as a \myred{single embedding by item tower}, which then uses the InfoNCE loss to fit the data distribution. 
    During inference, these multiple user embeddings are used to retrieve the other nearest neighbor items in the embedding space. E.g., Assigning an equal quota (e.g., 100) to each interest head/embedding, e.g, Kuaiformer~\cite{kuaiformer}, MIND~\cite{li2019multi}.
    \item Auto-regressive-based framework: inspired by the LLM, this branch of methods first tokenizes each item into a multi-level semantic IDs, and then utilize the users' historical sequence as input to generate next item multi-level Semantic ID in a regressive training manner.
    In inference, we could employ the beam search technique to generate hundreds of multi-level Semantic IDs to decode corresponding item candidates, e.g, TIGER~\cite{rajput2023recommender}.
\end{itemize}
Actually, both architectures contribute substantial progress to the RecSys community, while the ANN paradigm~\cite{andoni2018approximate} has supported many early traditional retrieval methods. The auto-regressive beam-search paradigm, aligning with LLM beam-search inference technique, opening a new direction for the generative retrieval.

\begin{figure}[t!]
  \centering
  \includegraphics[width=9cm,height=5cm]{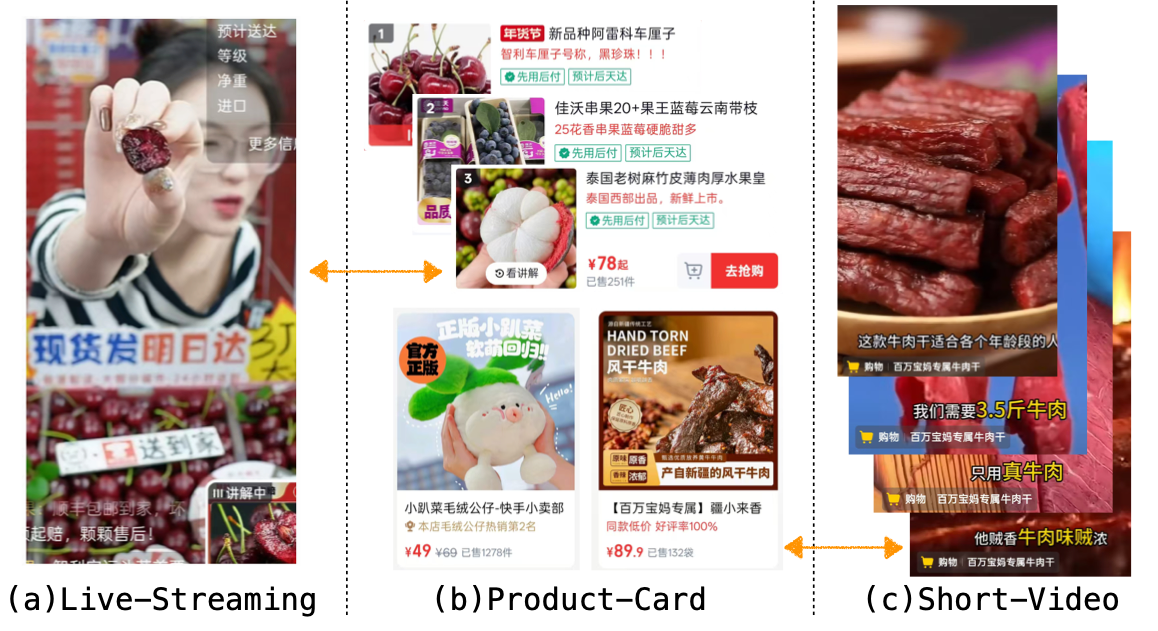}
  \caption{Three major shopping item categories at Kuaishou, where the product-card represents one specific product, the short-video typically connects to only one product-card, live-streaming can sell multiple product-cards simultaneously.}
  \label{itemcate}
  \vspace{-0.8cm}
\end{figure}

\myorange{\textbf{Motivation}}.
Inspired by the scaling-friendly auto-regressive paradigm TIGER, many researchers have started to deploy it at real industrial business services, to explore the GPT-moment in RecSys.
The most influential effort is the OneRec series~\cite{deng2025onerec, zhou2025onerec}, which provides a unified end-to-end generative recommendation framework aligned with advanced LLM technique:
\begin{itemize}
    \item \textit{Employing \myblue{pure Transformer architecture (pre-training)} backbone to generate next items candidates with beam search technique}.
    \item \textit{Applying the \myblue{RL (post-training)} to connect the retrieval base model and ranking reward model as an end-to-end system}.
\end{itemize}
While OneRec has preliminarily validated the feasibility of leveraging LLM-derived generative paradigms in RecSys, its design is inherently tailored to entertainment short-video service.
Nevertheless, for revenue e-commerce services, there are many incompatible demands with entertainment short-video services, including:
\begin{itemize}
    \item \textit{Instead of the Short-video, the E-commerce service has diverse item types such as product-cards, short-videos, live-streamings (as shown in Figure~\ref{itemcate}), it is required to consider inherent selling products attributes and content watching experience at same time.}
    \item \textit{In short-video services, users have a higher tolerance for watching various content and left enough positive behaviours, while in E-commerce content that users have higher decision-making costs, relying a long funnel of exposure to click to conversion, thus the positive behaviors are always extremely sparse}.
    \item \textit{Different from multiple objective optimization, e-commerce focuses on improving the Gross Merchandise Value (GMV) objective only, this singular objective imposes significantly higher demands on the model's capabilities}.
\end{itemize}
These issues have not been adequately addressed in previous works and need to be further explored.

\myorange{\textbf{Contribution}}.
To this end, we present our first-hand practical experience in e-commerce, and propose an end-to-end generative recommendation solution family for various item categories, termed as \textbf{OneMall}. 
In OneMall, we deeply investigated (i) the adaptation of item semantic tokenizers for e-commerce, and (ii) the generative model architectures to fuse long/short user interests and item attributes, and (iii) the design of reinforcement learning paradigms to connect reward ranking model, for richer positive behavior feedback and accuracy optimization direction.
Our OneMall key modification insights are as follows:
\begin{itemize}
    \item \textbf{Item Tokenizers for e-commerce}: As the foundation of generative recommendation, how to tokenize an item as multiple level Semantic IDs is extremely important, which should reflect both real-world item relations and business item relations simultaneously.
    Specifically, different item categories play different roles in our system: (1) the product-card items are only exposed to users at the Shop Tab, serving only shopping purposes; (2) the short-video and live-streaming items are exposed at both the Shop Tab and Entertainment Tab, and should consider both shopping and user-watch-experience simultaneously; (3) The live-streaming side Semantic IDs should be dynamically updated along with selling product changes.
    \item \textbf{Generative Model Architecture}: At the model designing stage, we mainly employ several Transformer variants as the backbone architecture, including the (1) Query Transformers~\cite{li2023blip} for longer sequence compression, (2) Cross Transformers for information extraction, and (3) Decoder-Style Sparse Mixture-of-Experts Transformers~\cite{liu2024deepseek} for the auto-regressive sequence generation. Meanwhile, to alleviate the semantic IDs conflict, we also conduct an in-batch contrastive learning with real target items as auxiliary objective, which significantly enhances our model prediction performance.
    \item \textbf{Reinforcement Learning}: In NLP area, how to encourage LLM to generate the high-quality response that aligned with human preference is a vital topic. Similarly, from the RecSys perspective, how to encourage the retrieval model to generate items that interest users with higher CTR/CVR rates (i.e., reward) is also important to our system. To address the long-standing issue of isolation in traditional cascaded retrieval-ranking pipelines, we thereby introduce the online ranking model as a reward model, enabling fine-grained supervision to perceive the relative quality among different items in real time end-to-end manner.
\end{itemize}

The main contributions of our work are as follows:
\begin{itemize}
    \item We elaborate an end-to-end generative recommendation framework for e-commerce services, including (i) diverse product category Tokenizer design, (ii) pure Transformer design, (iii) RL supervised by ranking model rewards. To the best of our knowledge, this is the first work that aligned with up-to-date NLP technology for e-commerce services, which will shed light on other researchers to explore a more robust e-commerce.
    \item We conduct extensive offline and online experiments at Kuaishou e-commerce service. The offline experiments show that all prediction tasks get significant improvements, and the online experiments +14.7\% GMV in Product-Card, +10.3\% GMV in Short-Video, and +4.9\% GMV in Live-Streaming.
    \item Our OneMall has been widely deployed on various services, supporting 400 million active users daily.
\end{itemize}

\section{Preliminary}

\begin{figure}[t!]
  \centering
  \includegraphics[width=8cm,height=3.5cm]{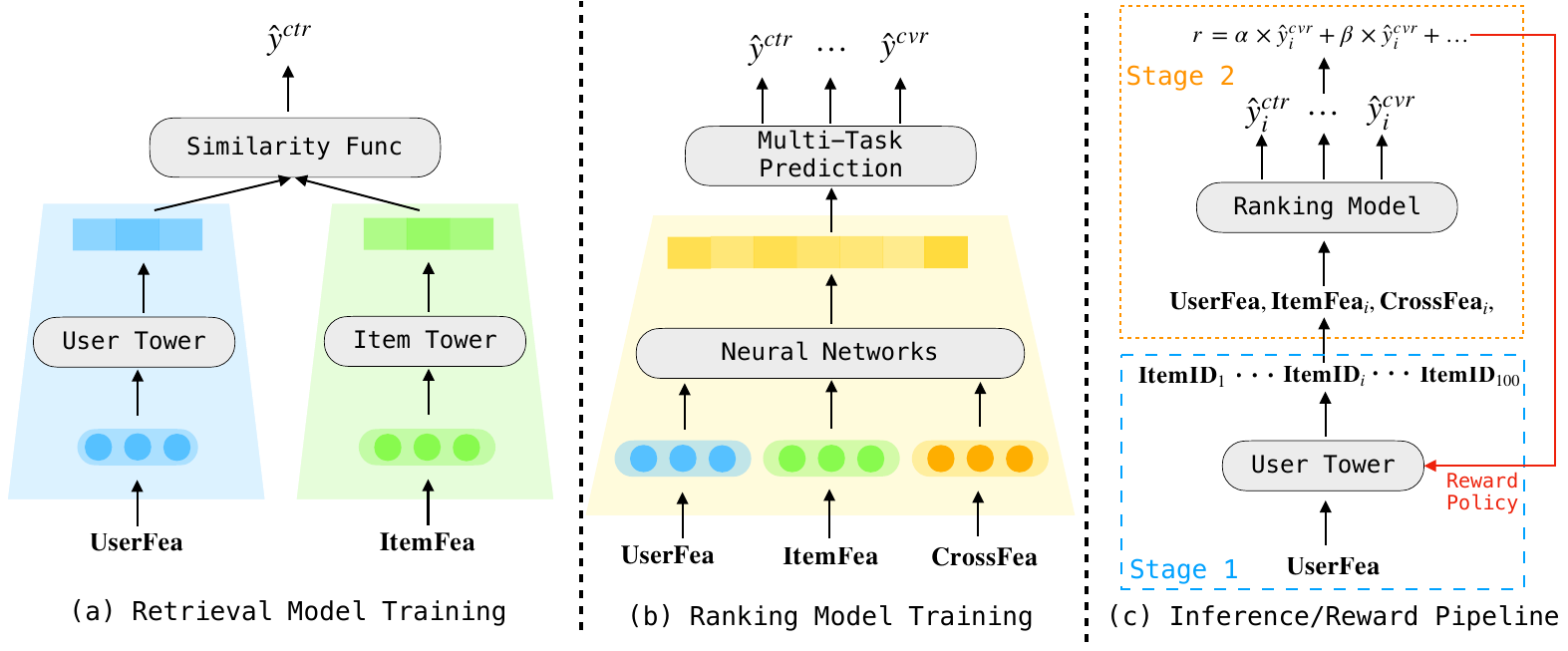}
  \caption{The model structure in industrial RecSys chain: (a) two-tower retrieval model; (b) more feature ranking model; (c) RecSys chain, first pass the user tower of retrieval model, and then feed to ranking model to get a fusion score.}
  \label{cascading}
  \vspace{-0.4cm}
\end{figure}

\subsection{Feature, Retrieval and Ranking Pipeline}
\label{chain}
Figure~\ref{cascading} illustrates a naive prototype of industry RecSys chain.
It consists of three basic elements: the Feature Engineer, Retrieval Model and Ranking Model:
\begin{enumerate}[leftmargin=*,align=left]
    \item \textbf{Feature Engineer}: for each user-item exposure at our application, it will be assembled features and real labels as a sample to feed to our model in one-epoch training setting~\cite{zhang2022towards}. Actually, those features can be divided into three groups: (i) \textit{user features}, e.g., latest clicked/buy sequences; (ii) \textit{target item features}, e.g., item tag, Semantic IDs; (iii) \textit{user-item cross features}, e.g., latest clicked/buy sequences with same target item tag.
    \item \textbf{Retrieval Model}: Although there exists three sets features, however, only the user and item set of features Retrieval models can used, since we can only utilize the user feature/component in RecSys chain (in Figure~\ref{cascading}(c)). Therefore, Retrieval model always follow Two-Tower paradigm with a single supervised objective, and skip user-item cross feature to avoid information leakage in Figure~\ref{cascading}(a).
    \item \textbf{Ranking Model}: In ranking model designing, to ensure the most precise prediction, all the features and labels will involved to support model training (as shown in Figure~\ref{cascading}(b)), hence the ranking model's prediction accuracy is significantly higher than retrieval model, and those prediction score will fuse as one joint score to sort item candidates, e.g., Pantheon enemble sort~\cite{cao2025pantheon}.
\end{enumerate}

\begin{figure*}[t!]
  \centering
  \includegraphics[width=18cm,height=4.5cm]{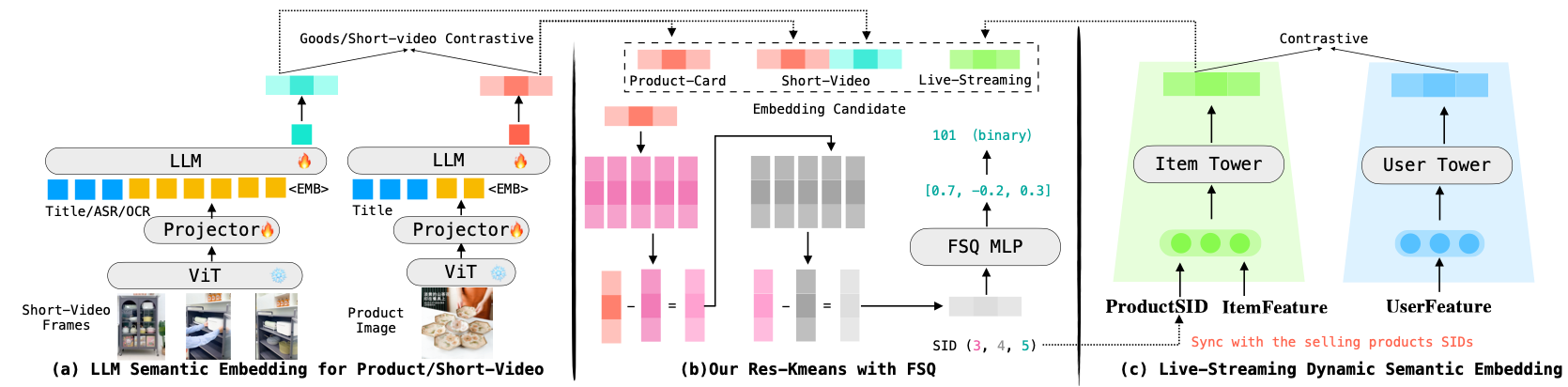}
  \caption{(a) LLM fine-tuning process in Product-Card/Short-Video Item2Item dataset; (c) Live-Streaming embedding generation with selling product Semantic IDs; (b) ResKmeansFSQ Tokenizer, different scenarios utilize different embedding candidates.}
  \label{semantic}
  \vspace{-0.4cm}
\end{figure*}

\subsection{Res-Kmeans Semantic Tokenizer}
\label{reskmeans}
Tokenizer is the cornerstone of auto-regressive token prediction paradigm, it aims at utilizing token ID to represent a specific unique semantic~\cite{song2021fast}.
In RecSys, how to transform an item of RecSys to a token sequence (i.e., Semantic IDs) is an opening topic, there are many works explored learnable/heuristic parallel/residual techniques in recent~\cite{luo2024qarm,shi2025llada}.
In OneMall, we largely apply the QARM's Res-Kmeans heuristic residual technique as Tokenizer to mapping an item embedding into multi-level Semantic IDs.
Particularly, the Res-Kmeans method has the following steps:
\begin{enumerate}[leftmargin=*,align=left]
\item \textit{Data collection}: Randomly sampled tens of millions of item embeddings set $\mathbf{M}\in \mathbb{R}^{N\times d}$, where the item embeddings could generate by finetuned-LLM or Item-Tower of Two-tower retrieval model.
\item \textit{Residual codebook training}: According to item embedding set $\mathbf{M}$, next conduct several layer \texttt{Kmeans} algorithm to obtain residual codebooks $\{\mathbf{C}^1, \mathbf{C}^2, \dots, \mathbf{C}^L\}$, where $\mathbf{C}^\cdot \in\mathbb{R}^{K\times d}$.
\begin{equation}
    \begin{split}
    \mathbf{C}^1 &= \texttt{Kmeans}(\mathbf{M}, K), \quad \mathbf{M}^1 = \mathbf{M}-\texttt{NearestRep}(\mathbf{M}, \mathbf{C}^1)\\
    \mathbf{C}^2 &= \texttt{Kmeans}(\mathbf{M}^1, K), \quad \mathbf{M}^2 = \mathbf{M}^1-\texttt{NearestRep}(\mathbf{M}^1, \mathbf{C}^2)\\
    &\dots,\quad \mathbf{C}^L = \texttt{Kmeans}(\mathbf{M}^{L-1}, K)\\
    \end{split}
    \label{rqcodebooktrain}
\end{equation} 
where the \texttt{NearestRep} denotes the nearest representation of corresponding codebook.
\item \textit{Semantic ID inference}: Once the residual codebooks $\{\mathbf{C}^1, \dots, \mathbf{C}^L\}$ is trained, it will be freezed to inference the existing and new-upload items to produce their Semantic IDs. For each item embedding \textbf{m}, the workflow as follows:
\begin{equation}
    \begin{split}
    c_1 &= \texttt{NearestCode}(\mathbf{m}, \mathbf{C}^1), \quad \mathbf{m}^1 = \mathbf{m}-\mathbf{C}^1_{c_1}\\
    c_2 &= \texttt{NearestCode}(\mathbf{m}^1, \mathbf{C}^2), \quad \mathbf{m}^2 = \mathbf{m}^1-\mathbf{C}^1_{c_2}\\
    &\dots,\quad c_L  = \texttt{NearestCode}(\mathbf{m}^{L-1}, \mathbf{C}^L)\\
    \end{split}
    \label{rqcodebookinfer}
    \end{equation}
\end{enumerate}
where the \texttt{NearestCode} denotes the nearest representation index of corresponding codebook, and its multi-level Semantic IDs can be formed as $\{c_1, c_2, \dots, c_L\}$.

\section{OneMall Workflow}
This Section express our OneMall designing insights for e-commerce.

\subsection{Semantic Tokenizer for E-commerce}
In e-commerce, each item is a distinct physical entity with specific semantic meaning, thus a fundamental requirement for Tokenizer is that it must be capable of preserving and explicitly \myred{general world Semantic}.
Meanwhile, for business perspective, e-commerce items typically come in three different categories: product-cards, short-videos, live-streamings in Figure~\ref{itemcate}.
Those follow different recommendation logics: (i) the product cards are only exposed when users enter the Shop Tab, which indicates users have a purchase intention.
(ii) For the short-videos and live streamings, however, are exposed across multiple channels (Shop Tab, Entertainment Tab, etc.).
Therefore, it is necessary to consider both the user’s watching experience and whether the products being sold are of interest to them. 
\myred{Further, the short-video typically sells only one item, but live streaming can sell multiple different items simultaneously}.
(iii) The live streaming's selling products may change over time, and the underlying semantics behind them should also evolve accordingly.
How to reflect the \myred{general real world item semantic} relation
and its business purpose, is the key point to build a robust Tokenizer for e-commerce, e.g., \myred{product-card (Commercial), Short-video (Commercial\&Watching), Live Streaming (Commercial\&Watching\&Dynamic)}.

To reach those limitations, we choose to fine-tune LLMs with business data as the Tokenizer, since LLM could provide fairness business knowledge and precise world semantic for all popular/long-tail items. 
Here we mainly extend QARM's Item2Item contrastive tuning pipeline~\cite{luo2024qarm}: (1) Business Item2Item Data Filter, (2) Semantic Compression Backbone, (3) Training/Inference Pipeline.

\subsubsection{Business Item2Item Data Filter}
Specifically, we collect two types of Item2Item data to consider the product intrinsic Commercial and Watching value.
\begin{itemize}
    \item \textit{Product-Card Intrinsic Commercial Relation}: To capture the business product intrinsic commercial relation, we export a large volume Item2Item pair with higher similarity within product-card items, e.g., nearest other items in two-tower retrieval model item embedding space, or Swing Item2Item statistics algorithm. According to the (product-card ID, product-card ID) corpus, we apply a naive downsample filter to ensure each item appears no more than 40 times to avoid exposure bias, and finally collect 70 million samples.
    \item \textit{Short-Video Watching Relation}: In order to inject the watching experience into the tokenizer, we further collect a corpus of entertainment short‑video and e‑commerce product‑card pairs.
    Specifically, for each positively clicked product item, we select the latest entertainment short‑videos that were long‑viewed with interval more than 1 minute and less than 5 minutes as a positive (short-video ID, product-card ID).
    Note that the short-video ID includes entertainment short-video and e-commerce short-video.
    To ensure the quality of the constructed dataset, we further implement three filtering strategies: (1) filtering `News', `Comedy' `Dance' `Film/TV' and `Selfie' Entertainment Short-video, (2) down-sample high-frequency products, (3) down-sample the Item2Item pair from same user, and collect 12 million samples.
\end{itemize}

\subsubsection{Semantic Compression Backbone}
Based on the filtered Commercial (product-card ID, product-card ID) and Watching (short-video ID, product-card ID) Item2Item corpus, we next utilize them as alignment supervision signal to fine-tune an LLM to adapt e-commerce business.
For the LLM inputs, we only consider the Product and Short-Video information:
\begin{itemize}
    \item \textit{Product}: The main image (224*224) and its title.
    \item \textit{Short-Video}: Sample 6 image (224*224) and its title/OCR/ASR.
\end{itemize}
Based on them, we apply the Swin-Transformer as vision encoder and Qwen2.5 1.5B as Text encoder, the organization of those visual and text tokens are shown in the Figure~\ref{semantic}(a).
Note that we freeze the ViT and fine-tune the Projector and LLM a with InfoNCE contrastive objective, and the last hidden state of the special token \texttt{<EMB>} served as the final representation.

\subsubsection{Semantic ID Generation Pipeline}
Up to now, we have introduced how to train an LLM to generate the semantic embedding $\mathbf{m}$ that with general real world semantic and business knowledge.
To transform them as Semantic IDs, we devise an elaborate pipeline for the Product-card, Short-Video and Live-streaming inference, which includes two quantization techniques: Res-Kmeans and FSQ~\cite{mentzer2023finite}.
For the first two layers' Res-Kmeans, we run it as follows:
\begin{itemize}
    \item \textit{Product-card}: for the product-card, since it only represent commerce intent, thus we only use the LLM generated product-card embedding as $\mathbf{m}$ to quantify.
    \item \textit{Short-video}: for the e-commerce short-video, it needs to reflect both commerce and watching experience simultaneously, thus we concatenate the LLM generated product-card embedding and the short-video embedding as $\mathbf{m}$ to quantify.
    \item \textit{Live-streaming}: for live-streaming, its content and selling items dynamically change, thus we could not generate its Semantic ID directly.
    In OneMall, we utilize the two-tower style retrieval model's item tower generated embedding to quantify, and the real-time selling products' Semantic IDs served as additional feature. Moreover, to alleviate the excessive change rate of semantic codes, we used a lower learning rate for the item tower.
\end{itemize}
For the last layer FSQ~\cite{mentzer2023finite}, we train a binary 16-bit MLP to quantify the Residual embedding $\mathbf{m}-\mathbf{C}^1_{c_1} - \mathbf{C}^2_{c_2}$ as a 4096 code, it can significantly reduce the conflict rate.
For the sake of readability, we do not elaborate on the specific formulas here; an overview of the entire process is illustrated in the accompanying in Figure~\ref{semantic}(b)(c).

\begin{figure}[t!]
  \centering
  \includegraphics[width=7cm,height=7cm]{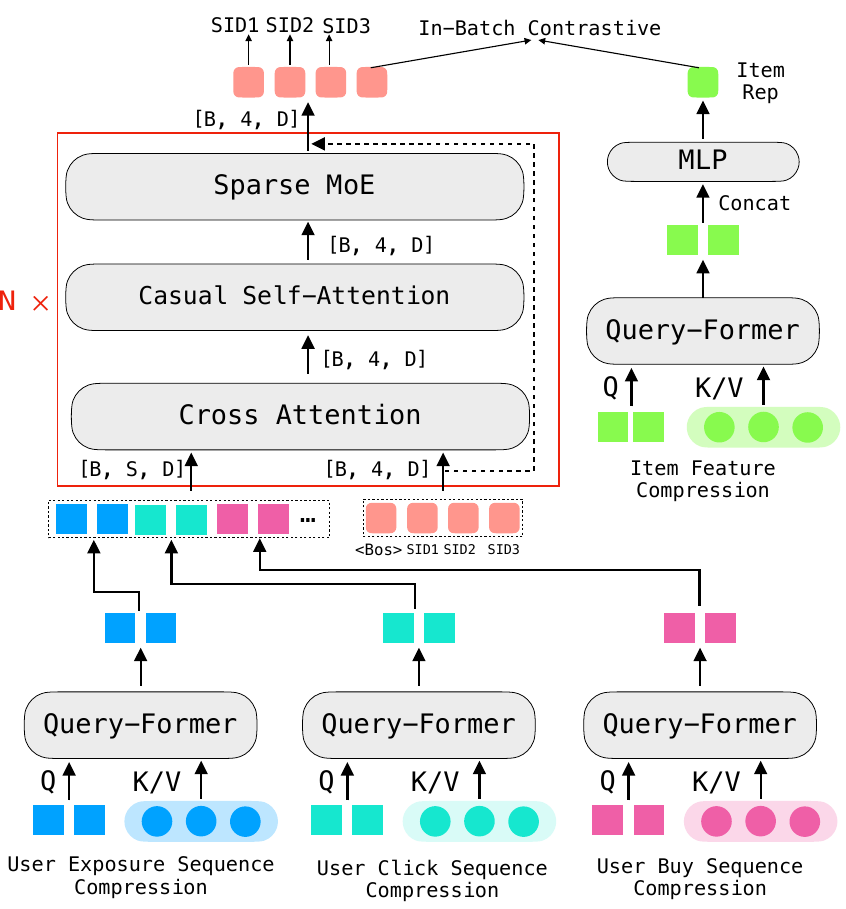}
  \caption{The Transformer backbone of OneMall.}
  \label{backbone}
  \vspace{-0.4cm}
\end{figure}

\subsection{Decoder-Only Model Architecture}
At OneMall, we employ several Transformer variants served as the backbone architecture, including the Query-Former~\cite{li2023blip}, Cross-Attention, Casual-Self-Attention, as shown in Figure~\ref{backbone}.
Here we list the designing insights as follows:
\begin{itemize}
    \item \textit{Query Transformers for Longer Sequence Compression}: In e-commerce service, users always have higher decision-making costs, relying a long funnel of exposure to click to buy an item.
    As a result, our OneMall needs to utilize a variety of different behavioral sequences to enrich user's preference from different aspects, such as Exposure/Click/Buy sequences and so on.
    In sequence modeling, if those sequences are concatenated into one, it becomes too long to afford the complexity of Transformer.
    To alleviate it, we first apply the Query-Former technique to compress the longer sequence into dozens continuous representation:
    \begin{equation}
    \footnotesize
    \begin{split}
    \mathbf{F}_{\texttt{click}} = \texttt{QFormer}_{\texttt{click}}(\mathbf{Q}_{\texttt{click}}, \mathbf{H}_{\texttt{click}}, \mathbf{H}_{\texttt{click}})
    \end{split}
    \label{queryformer}
    \end{equation}
    where the $\mathbf{F}_{\texttt{click}}\in \mathbb{R}^{M\times D}, \mathbf{Q}_{\texttt{click}}\in \mathbb{R}^{M\times D}, \mathbf{H}_{\texttt{click}}\in \mathbb{R}^{H\times D}$, $M$ denotes the query token number (e.g., 10), $H$ indicates the input sequence length (e.g., 500), $D$ is the embedding dimension (e.g., 128). 
    In this way, our OneMall could consider multiple different sequences with a lower computation cost, such as $\mathbf{F}_{\texttt{Buy}}$, $\mathbf{F}_{\texttt{Exposure}}$.
    Besides, we also utilize the Query-Former to compress item side features to generate its representations:
    \begin{equation}
    \footnotesize
    \begin{split}
    \mathbf{f}_{\texttt{item}} = \texttt{MLP}(\texttt{Flatten}(\mathbf{F}_{\texttt{item}})), \ \ 
    \mathbf{F}_{\texttt{item}} = \texttt{QFormer}_{\texttt{item}}(\mathbf{Q}_{\texttt{item}}, \mathbf{I}_{\texttt{item}}, \mathbf{I}_{\texttt{item}})
    \end{split}
    \label{itemqueryformer}
    \end{equation}
    where the $\mathbf{f}_{\texttt{item}}\in \mathbb{R}^D$ is the final item embedding, $\texttt{Flatten}(\cdot)$ is a naive operation to flatten model inputs, $\mathbf{I}_{\texttt{item}}$ is the item-side feature sequence.
    \item \textit{Cross Transformers for historical information extraction}: Next, to fuse the historical user side information $\{\mathbf{F}_{\texttt{click}}, \mathbf{F}_{\texttt{exposure}}, \mathbf{F}_{\texttt{Buy}, \dots}\}$ to predict next item Semantic IDs $\{s_1, s_2, s_3\}$, we further employ the Cross Attention mechanism for the low latency extraction:
    \begin{equation}
    \footnotesize
    \begin{split}
    \{\hat{\mathbf{s}}_0^L,\hat{\mathbf{s}}_1^L,\hat{\mathbf{s}}_2^L,\hat{\mathbf{s}}_3^L\} = \texttt{Cross-Att}^L(&\{\mathbf{s}_0^{L-1},\mathbf{s}_1^{L-1},\mathbf{s}_2^{L-1},\mathbf{s}_3^{L-1}\}\cdot\mathbf{W}_q^L,\\
    &\{\mathbf{F}_{\texttt{click}}, \mathbf{F}_{\texttt{exposure}}, \mathbf{F}_{\texttt{buy}}, \dots\}\cdot\mathbf{W}_k^L,\\
    &\{\mathbf{F}_{\texttt{click}}, \mathbf{F}_{\texttt{exposure}}, \mathbf{F}_{\texttt{buy}}, \dots\}\cdot\mathbf{W}_v^L)
    \end{split}
    \label{crossattention}
    \end{equation}
    where the $\mathbf{W}_q^L/\mathbf{W}_k^L/\mathbf{W}_v^L\in \mathbb{R}^{D\times D}$ are the learnable parameter matrices of $L$-th layer Cross-Attention, $\{\mathbf{s}_0^{L-1},\mathbf{s}_1^{L-1},\mathbf{s}_2^{L-1},\mathbf{s}_3^{L-1}\}$ denotes the corresponding inputs.
    Note that the $\{\mathbf{s}_0^0,\mathbf{s}_1^0,\mathbf{s}_2^0,\mathbf{s}_3^0\}$ is the LookUp embedding of $\{\texttt{<BOS>},s_1, s_2, s_3\}$.
    \item \textit{Decoder-Style Sparse MoE Transformers for auto-regressive sequence generation}: To further scale the model capacity while maintaining efficient inference, we adopt a Decoder-Style Sparse Mixture-of-Experts (MoE) Transformer~\cite{liu2024deepseek} as the core of the auto-regressive generation module.
    For each layer of the decoder, we first apply causal self-attention to the Semantic ID sequence to preserve auto-regressive constraints, then pass the output through the sparse MoE to capture complex semantic patterns in Semantic IDs:
    \begin{equation}
    \footnotesize
    \begin{split}
    \{\bar{\mathbf{s}}_0^L, \bar{\mathbf{s}}_1^L&, \bar{\mathbf{s}}_2^L, \bar{\mathbf{s}}_3^L\} = \texttt{Casual-Self-Att}(\{\hat{\mathbf{s}}_0^L, \hat{\mathbf{s}}_1^L, \hat{\mathbf{s}}_2^L, \hat{\mathbf{s}}_3^L\}),\\
    \{\mathbf{s}_0^L, \mathbf{s}_1^L, \mathbf{s}_2^L, \mathbf{s}_3^L\} = &\texttt{Sparse-MoE}(\{\bar{\mathbf{s}}_0^L, \bar{\mathbf{s}}_1^L, \bar{\mathbf{s}}_2^L, \bar{\mathbf{s}}_3^L\}) + \{\mathbf{s}_0^{L-1}, \mathbf{s}_1^{L-1}, \mathbf{s}_2^{L-1}, \mathbf{s}_3^{L-1}\}
    \end{split}
    \label{selfattention}
    \end{equation}
    where the $\texttt{Casual-Self-Att}(\cdot)$ indicates the standard masked self-attention, and the $\texttt{Sparse-MoE}(\cdot)$ is the key component to scaling model parameter. For the expert load-balancing, we use the loss-free mechanism to constrain it.

    \item \textit{Supervised Objectives}: For the model supervision signals, we consider the auto-regressive next item's Semantic IDs generation as main training objective and a two-tower style in-batch contrastive task as an auxiliary objective at same time:
    \begin{equation}
    \footnotesize
    \begin{split}
    \mathcal{L}_{\texttt{NTP}} = \texttt{So}&\texttt{ftmax}(\{\mathbf{s}_0^L, \mathbf{s}_1^L, \mathbf{s}_2^L\},\{s_1, s_2, s_3\}),\\
    \mathcal{L}_{\texttt{contrastive}} = &\texttt{In-batch-contrastive}(\mathbf{s}_3^L, \mathbf{f}_{\texttt{item}})\\
    \end{split}
    \label{supervision}
    \end{equation}
    where the $\mathcal{L}_{\texttt{NTP}}$ denotes the auto-regressive next-token prediction training objective, and the $\mathcal{L}_{\texttt{contrastive}}$ follows the traditional ANN methods in-batch contrastive settings.
    The difference is that: $\mathbf{s}_3^L$ already encodes the full target item's Semantic ID sequence, which results in very high accuracy (over 98\% in accuracy@1) for contrastive learning. Therefore, in online inference, we only employ beam search for retrieval.
\end{itemize}

\begin{figure}[t!]
  \centering
  \includegraphics[width=9cm,height=3.5cm]{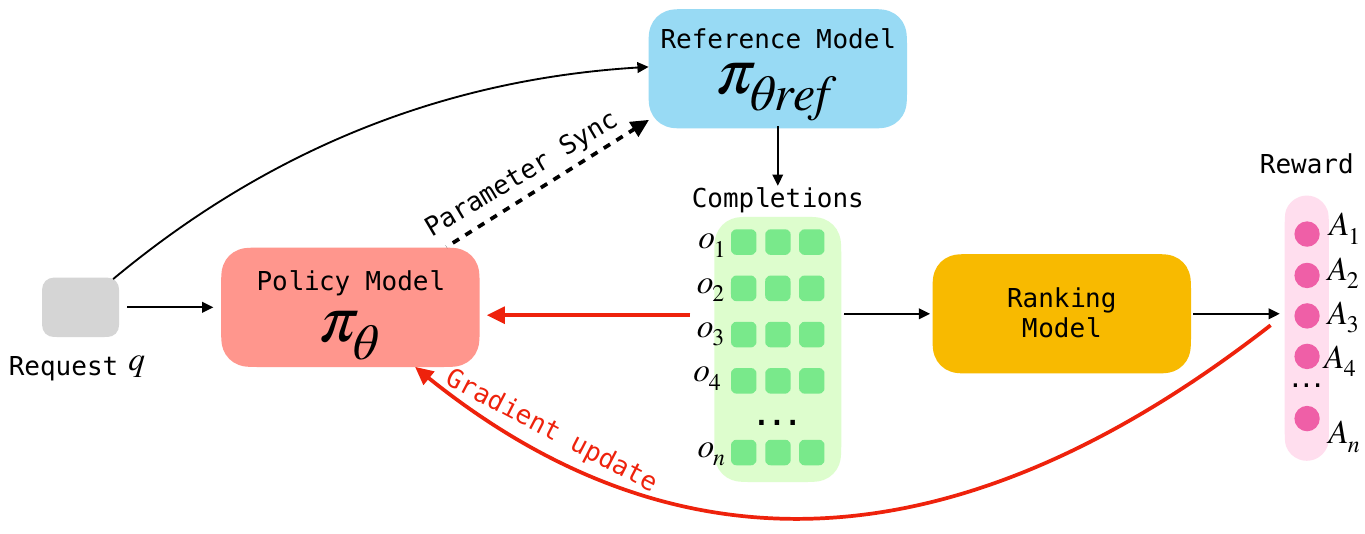}
  \caption{The RL workflow of our OneMall.}
  \label{rlflow}
  \vspace{-0.5cm}
\end{figure}

\begin{table*}[t!]
\centering
\caption{Parameter Scaling Validation at Scaling Performance at the E-commerce Short-video scenario.}
\setlength{\tabcolsep}{4pt}{

\begin{tabular}{c|ccccc|ccc|ccc}
\toprule
Parameter &Layer&Att-Dim&FFN-Dim&Head &Expert&Acc-SID1 &Acc-SID2 & Acc-SID3 &HR@50 & HR@100 &HR@500\\
\midrule
0.05B &2&1024&4096&8&- &14.5\% &43.9\% &61.0\% &32.9\% &41.3\%&60.5\%\\
0.1B&4&1024&4096&8&-&14.9\% &46.2\% &63.6\% &34.7\%&43.8\%&63.2\%\\
0.5B-A0.1B&4&1024&4096&8&12/2&16.0\% &50.6\% &69.3\% &44.7\%&56.0\%&74.2\%\\
1.3B-A0.1B&4&1024&4096&8&24/2&16.2\% &51.7\% &71.7\% &45.6\%&57.3\%&76.0\%\\
\bottomrule
\end{tabular}
}
\label{scaling}
% }
\end{table*}

\subsection{Reinforcement Learning Policy}
From the early PPO/DPO (DPO is an approximate reinforcement learning approach) to the later GRPO/GSPO~\cite{rafailov2023direct, guo2025deepseek, zheng2025group}, the RL driven post-training paradigm provides a new impressive data scaling direction to enhance LLM understanding and reasoning ability~\cite{yang2025qwen3}.
% %
%
To fully unleash the potential of reinforcement learning and align with advanced LLM post-training techniques in RecSys, we further design a retrieval-ranking connection pipeline. This enables OneMall to generate items that better match user interests and achieve higher rewards.
Formally, we take the online ranking model (could utilize all the user/item/cross features and interaction labels) as a reward model, thereby it can provide more fine-grained prediction probabilities from different aspects, e.g., CTR/CVR/CTCVR/EGPM and others.
Based on those multiple prediction probabilities, we first apply a simple F1 scoring formula to fuse them to estimate the reward score $r$:
    \begin{equation}
    \footnotesize
    \begin{split}
    r = \alpha\times \hat{y}_{\texttt{ctr}} + \beta\times \hat{y}_{\texttt{ctcvr}} + \gamma\times \hat{y}_{\texttt{egpm}} + \dots
    \end{split}
    \label{ensemble}
    \end{equation}
where the $\hat{y}_{\texttt{ctr}}/\hat{y}_{\texttt{ctcvr}}/\hat{y}_{\texttt{egpm}}$ are the ranking model outputs, $\alpha$, $\beta$, $\gamma$ are the hand-craft factors, which reflect the e-commerce worth of each user-item pair.
For notation brevity, our RL pipeline can be described as follows:
\begin{itemize}
    \item \textit{Policy model} $\pi_\theta$: the NTP model parameter in streaming training.
    \item \textit{Reference model} $\pi_{\theta ref}$: parameters are synchronized from the policy model periodically, for each query $q$, the $\pi_{\theta ref}$ could sample a set of response \myred{$\left\{o_{1}, o_{2}, \cdots, o_{n}\right\}$}.
    \item \textit{Reward model}: according to those responses \myred{$\left\{o_{1}, o_{2}, \cdots, o_{n}\right\}$}, the reward model further normalized as the advantage scores for each response as \myred{$\left\{A_{1}, A_{2}, \cdots, A_{n}\right\}$ ($A_i = \frac{r_i - \texttt{mean}(r)}{\texttt{std}(r)}$)}, to supervise which responses are encouraged or need to punish.
    \item \textit{Optimize Strategy}: regarding to the responses and their advantages, we can utilize different training strategies to optimize the Policy model parameters $\pi_\theta$.
\end{itemize}
The overall training sketches is shown in Figure~\ref{rlflow}, while we first employing the Reference model to produce completions, then the ranking model to generate corresponding the Reward advantages, and finally to optimize our base policy retrieval model in end-to-end manner, breaking the barrier of retrieval and ranking model.
We sample 2\% offline training samples as the simulate request $q$ to run the RL policy, and have attempted the DPO/GRPO policy to inject the e-commerce reward knowledge to optimize our OneMall:
\begin{itemize}
\item \textit{DPO}: We rank the sampled responses according to their advantage scores $A$, treating the highest-scoring response as the positive Semantic IDs $o_{pos}$ and the lower-scoring ones as negative Semantic IDs $o_{neg}$, which are then used to supervise the optimization of the policy model.
    \begin{equation}
    \footnotesize
    \begin{split}
    \mathcal{L}_{\texttt{DPO}} = - \sigma(\lambda \log \frac{\pi_\theta (o_{pos}|q)}{\pi_{\theta ref}(o_{pos}|q)+\delta} - \lambda \log \frac{\pi_\theta (o_{neg}|q)}{\pi_{\theta ref}(o_{neg}|q)+\delta})
    \end{split}
    \label{dpo}
    \end{equation}
    where the $\lambda$ is a hand-craft hyper-coefficient, e.g., 0.1-0.5, $\delta$ is a small to avoid zero division.
    \item \textit{GRPO}: We randomly sample $m$ completions from $n$ candidates:
    \begin{equation}
    \footnotesize
    \begin{split}
    \mathcal{L}_{\texttt{GRPO}} = - \sum_{i}^{m}\min(\texttt{clip}(\frac{\pi_\theta (o_{i}|q)}{\pi_{\theta ref}(o_{i}|q)+\delta},1-\epsilon,1+\epsilon)\times A_i)
    \end{split}
    \label{grpo}
    \end{equation}
    where the $\texttt{clip}(\cdot)$ function truncates the gradients to prevent a few exceptionally samples from dominating the entire batch, thereby ensuring training stability.
\end{itemize}

Finally, the RL loss is jointly optimized together with the NTP loss and the two-tower contrastive learning loss in Eq.(\ref{supervision}), thereby injecting richer positive behavior feedback and accuracy optimization direction: $\mathcal{L} = 0.5\cdot\mathcal{L}_{\texttt{RL}} + \mathcal{L}_{\texttt{NTP}} +\mathcal{L}_{\texttt{contrastive}}$.

\section{Experiments}
In this Section, we conduct extensive offline and online experiments across Product-Card, Short-Video and Live-Streamings e-commerce services.
For a fair performance estimation, we apply three metrics to compare with baselines and our model variants:
\begin{itemize}
    \item \textit{Offline SID Accuracy}: The Accuracy metrics are used to measure the offline performance of same series models variants, here we shown the Accuracy@1 at three-level SID NTP precision.
    \item \textit{Simulation Hit Rate}: For a fair comparison with different methods, we replay the online user request as a simulation environment to collect the hit rate between the real viewed items whether they are retrieved by the corresponding models, e.g., does the viewed item is retrieved in the top 50/100/500 results.
    \item \textit{Online A/B Metrics}: These metrics evaluate the real-world performance of our System through controlled online experiments. Key indicators include Orders, Gross Merchandise Value (GMV), Click-Through Rate (CTR), Conversion Rate (CVR), and Gross Profit Margin (GPM).
\end{itemize}

\begin{figure}[t!]
  \centering
  \includegraphics[width=9cm,height=2.8cm]{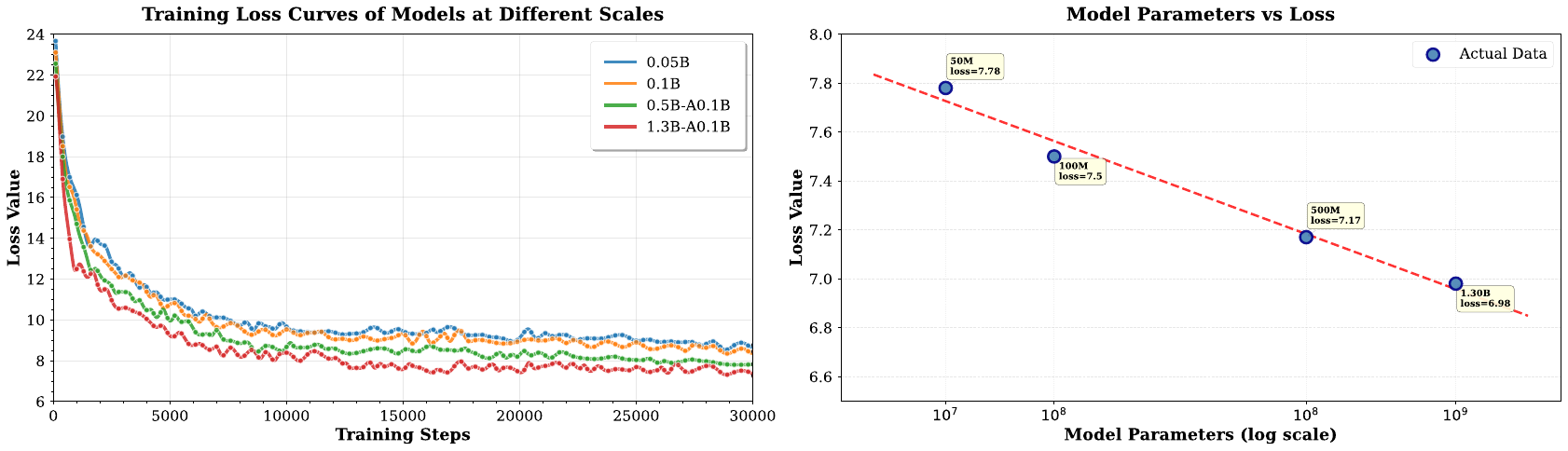}
  \caption{OneMall scaling trend at the Short-video scenario.}
  \label{scalingpara}
  \vspace{-0.6cm}
\end{figure}

\subsection{Offline Scaling Experiments}
As a generative recommendation paradigm, we first validate the scaling behavior of OneMall, with results shown in Table~\ref{scaling} and Figure~\ref{scalingpara}.
Specifically, we explore model scaling from 0.05B to 1.3B total parameters. However, due to computational resource constraints in online serving, we limit the maximum activated parameters to 0.1B, while scaling the remaining parameters through Sparse Mixture-of-Experts (MoE) with loss-free load balancing.

From Table~\ref{scaling}, we observe consistent improvements across all metrics as model capacity increases:
(1) \textit{SID Accuracy}: The three-level Semantic ID prediction accuracy (Acc-SID1/2/3) steadily improves from 14.5\%/43.9\%/61.0\% at 0.05B to 16.2\%/51.7\%/71.7\% at 1.3B-A0.1B, demonstrating that larger models better capture the hierarchical semantic structure.
(2) \textit{Hit Rate}: The simulation hit rate metrics (HR@50/100/500) show significant gains, improving from 32.9\%/41.3\%/60.5\% to 45.6\%/57.3\%/76.0\%, indicating enhanced retrieval quality.
Notably, the transition from dense 0.1B to sparse 0.5B-A0.1B yields the most substantial improvement (+10\% HR@50), suggesting that MoE architecture effectively expands model capacity while maintaining inference efficiency.

\begin{table}[t!]
% \footnotesize
\centering
\caption{Simulation Replay Performance Comparison.}
\setlength{\tabcolsep}{1pt}{
% \resizebox{\linewidth}{!}{

\begin{tabular}{c|c|ccc}
\toprule
Scenario & Parameter &HR@50 &HR@100 & HR@500 \\
\midrule
\multirow{3}{*}{\cellcolor{white}\textbf{Product-Card}} &SASRec-0.5B-A0.1B &23.2\%&30.4\%&45.1\% \\
&TIGER-0.5B-A0.1B &29.5\%&37.1\%&53.7\% \\
&OneMall-0.5B-A0.1B &34.3\%&40.9\%&59.3\% \\
\midrule
\multirow{3}{*}{\cellcolor{white}\textbf{Short-Video}} &SASRec-0.5B-A0.1B &30.2\%&41.9\%&66.3\% \\
&TIGER-0.5B-A0.1B &35.1\%&47.8\%&72.5\% \\
&OneMall-0.5B-A0.1B &44.7\%&56.0\%&74.2\% \\
\midrule
\multirow{3}{*}{\cellcolor{white}\textbf{Live-Streaming}} &SASRec-0.5B-A0.1B &53.9\%&58.8\%&68.4\% \\
&TIGER-0.5B-A0.1B &57.3\%&63.5\%&73.7\% \\
&OneMall-0.5B-A0.1B &65.9\%&69.1\%&80.5\% \\
\bottomrule
\end{tabular}
}
\label{simulation}
% }
\vspace{-0.6cm}
\end{table}

\subsection{Simulation Experiments}
We deploy the 0.5B-A0.1B version of OneMall across all e-commerce scenarios. To validate the effectiveness of our approach, we conduct fair comparisons on replayed online traffic, with results shown in Table~\ref{simulation}.
All compared methods use identical input features, but differ in their modeling paradigms:
(1) \textit{SASRec-0.5B-A0.1B}: A classic ANN-based retrieval method that compresses user interests into embeddings for nearest neighbor search.
(2) \textit{TIGER-0.5B-A0.1B}: An Encoder-Decoder generative retrieval method with a pre-trained RQ-VAE tokenizer strategy.
From Table~\ref{simulation}, we observe that OneMall consistently achieves the best performance across all scenarios:
These results validate the effectiveness of our e-commerce semantic tokenizer design, the Decoder-Only architecture, and RL.

\subsection{Online Experiments}
To validate the real business contribution that OneMall brings to our e-commerce services, we conducted online A/B testing experiments across Product-Card, Short-Video, and Live-Streaming scenarios.
We evaluate model performance using core e-commerce metrics, including Exposure, Click, Order, and Gross Merchandise Volume (GMV), as shown in Table~\ref{mainonlineshopping}.

From the results, OneMall achieves significant improvements across all scenarios:
(1) In \textit{Product-Card}, OneMall obtains +2.851\% Exposure, +13.01\% Click, +8.35\% Order, and +14.71\% GMV, demonstrating strong commercial value in the Mall Tab where users have explicit purchase intent.
(2) In \textit{Short-Video}, we observe +5.76\% Click and +11.65\% Order and 10.33\% GMV improvements, indicating that our tokenizer effectively captures both watching experience and commercial intent for e-commerce short-videos.
(3) In \textit{Live-Streaming}, OneMall achieves +4.47\% Order and +4.90\% GMV improvements, validating the effectiveness of our dynamic Semantic ID generation strategy for live-streaming content.
Overall, these online results confirm that OneMall successfully bridges the gap between user interest modeling and commercial conversion, bringing substantial business value to Kuaishou's e-commerce platform.

\begin{table}[t!]
% \footnotesize
\centering
\caption{Online A/B testing results of E-commerce services.}
\setlength{\tabcolsep}{3.5pt}{
% \resizebox{\linewidth}{!}{

\begin{tabular}{c|cccc}
\toprule
Scenarios&Exposure & Click &  Order  & GMV\\
\midrule
Product-Card &+2.851\% &+13.01\% &+8.35\% &+14.71\% \\
Short-Video&+15.32\% &+5.76\% &+11.65\% &+10.33\%\\
Live-Streaming&+2.78\% &+2.53\% &+4.47\% &+4.90\%\\
\bottomrule
\end{tabular}
}
\label{mainonlineshopping}
% }
\vspace{-0.4cm}
\end{table}

\begin{table}[t!]
% \footnotesize
\centering
\caption{RL Ablation Study.}
\setlength{\tabcolsep}{4pt}{
% \resizebox{\linewidth}{!}{

\begin{tabular}{c|ccccc}
\toprule
Beam Top &Strategy &  CTR   & CTCVR & GPM & Reward\\
\midrule
\multirow{3}{*}{\cellcolor{white}\textbf{Top10}} &Base &7.816\% &0.243\% &0.0983 & 0.2493\\
&DPO &7.834\% &0.254\% &0.0997 & 0.2543 \\
&GRPO &7.874\% &0.266\% &0.1020 & 0.2607\\
\midrule
\multirow{3}{*}{\cellcolor{white}\textbf{Top100}} &Base &4.369\% &0.106\% &0.0390 &0.1147\\
&DPO &4.396\% &0.110\% &0.0399 & 0.1169\\
&GRPO &4.403\% &0.112\% &0.03998 &0.1177 \\
\midrule
\multirow{3}{*}{\cellcolor{white}\textbf{Top500}} &Base &2.188\% &0.038\% &0.0127 &0.0463 \\
&DPO &2.210\% &0.039\% &0.0130 &0.0470 \\
&GRPO &2.225\% &0.041\% &0.0132 &0.0476 \\
\bottomrule
\end{tabular}
}
\label{rlabl}
% }
\vspace{-0.6cm}
\end{table}

\subsection{RL Effectiveness Analysis}
In this section, we thoroughly validate the effectiveness of reinforcement learning in e-commerce recommendation.
The core business objectives in e-commerce are CTR, CTCVR, and EGPM. We first perform score fusion as defined in Eq.(\ref{ensemble}), with $\alpha=1.0$, $\beta=30.0$, $\gamma=1.0$ to maintain consistent magnitude across different objectives.
We sample 2\% of training traffic for RL training with a beam size of 768 for candidate generation.
Additionally, we set the loss weight ratio ($\mathcal{L}_{\texttt{NTP}} + 0.5\mathcal{L}_{\texttt{RL}}$). We empirically find that excessively large RL loss weights degrade the SID accuracy metrics and affect the validity rate of generated candidates; thus we balance generation efficiency and quality with the hand-craft weight.
From Table~\ref{rlabl}, we have: 
(1) \textbf{DPO/GRPO vs. Base}: Both DPO and GRPO achieve consistent improvements over the Base model across all candidate segments (Top10/100/500). The reward term and score also improve at different degrees, demonstrating the effectiveness of RL in generating high-quality candidates and its capability to optimize multiple objectives simultaneously in recommendation.
(2) \textbf{GRPO vs. DPO}: GRPO outperforms DPO across all candidate segments, with particularly notable improvements in Top10 (e.g., +0.040\% CTR, +0.012\% CTCVR, +0.228\% GPM over DPO). The reason might be that GRPO normalizes rewards across all 768 sampled candidates to compute advantage scores, enabling the model to learn the quality distribution within each group. In contrast, DPO only uses pairwise positive-negative samples, providing less training feedback.

\begin{table}[t!]
% \footnotesize
\centering
\caption{Code Conflict Component Analysis.}
\setlength{\tabcolsep}{2pt}{
% \resizebox{\linewidth}{!}{

\begin{tabular}{l|cc|ccc}
\toprule
Variants-0.3B-A0.1B &Conflict & Exclusive &  HR@50   & HR@100 & HR@500 \\
\midrule
ResKmeans &36\% &86\% &33.9\% &43.0\% &62.6\% \\
ResKmeansFSQ &11\% &95\% &35.4\% &45.3\% &66.7\% \\
\midrule
$\quad$ +Aux Loss &11\% &95\% &+1.5\% &+1.7\% &+1.7\% \\
\bottomrule
\end{tabular}
}
\label{fsqabl}
% }
\vspace{-0.4cm}
\end{table}

\begin{table}[t!]
% \footnotesize
\centering
\caption{Query-Former Component Analysis.}
\setlength{\tabcolsep}{2.5pt}{
% \resizebox{\linewidth}{!}{

\begin{tabular}{l|cc|ccc}
\toprule
Variants &SeqLen & GFLOPs &  HR@50   & HR@100 & HR@500 \\
\midrule
QFormer &160 &9.2 &44.7\% &56.0\% &74.2\% \\
w/o QFormer &1205 &34.4 &+0.5\% &+0.3\% &+0.4\% \\
\bottomrule
\end{tabular}
}
\label{qformer}
% }
\vspace{-0.6cm}
\end{table}

\subsection{Component Analyses}
\textbf{Tokenizer Strategy}.
In our early version, we directly applied the vanilla three-layer 4096 Res-Kmeans codebook for model training, but encountered severe \textit{code collision} problem (i.e., one set of Semantic IDs maps to multiple items).
We hypothesize that this is because K-means optimization objective minimizes the average intra-cluster distance to centroids, without considering inter-cluster distances. This leads to cluster center collapse, resulting in high tokenizer collision rates and uneven code distribution.

To address this issue, we introduce FSQ (Finite Scalar Quantization) at the last quantization layer.
FSQ fixes the cluster centers in advance, ensuring uniform distribution of codes. Although this sacrifices some semantic information, it makes the partitioned space more regular.
As shown in Table~\ref{fsqabl}, ResKmeansFSQ combines the advantages of both approaches: it preserves semantic information from the first two Res-Kmeans layers while ensuring uniform cluster center distribution through FSQ.
\textbf{Auxiliary Contrastive Loss}.
Since Semantic IDs are relatively coarse-grained and lack item-specific fine-grained features, we introduce additional item-side information into the model.
Specifically, we feed important item attributes (e.g., product category, price, shop information) into an item tower and apply in-batch contrastive learning as an auxiliary objective.
Our experiments show that this auxiliary loss yields consistent improvements in hit rate (+1.5\%/+1.7\%/+1.7\% on HR@50/100/500), demonstrating the effectiveness of incorporating fine-grained item features to complement the coarse-grained Semantic ID representation.
\textbf{Query-Former Compression}.
To avoid the higher computation cost by long sequence, we introduce the Q-former for early sequence compression.
Fortunately, we find that its performance is on par with the longer sequence version, as shown in Table~\ref{qformer}, the GFLOPs (calculation cost per sample) is largely reduced with only minor performance degradation.

\section{Conclusion}
In this paper, we present OneMall, an end-to-end generative recommendation framework specifically designed for e-commerce services. 
By aligning with the successful paradigms of LLMs, OneMall integrates three key innovations: (1) unified semantic tokenizer solution, (2) Pure Transformer-based network designing, and (3) RL paradigm to distill ranking model knowledge.
Extensive experiments on Kuaishou's e-commerce platform demonstrate significant improvements across product cards, short-videos, and live-streaming scenarios.
We hope our practical experience will inspire future research on applying LLM techniques to industrial e-commerce recommendation.
In the future, we will explore: (1) unifying retrieval and ranking ability to OneMall, (2) text-based reasoning ability, (3) shared backbone for multi-service data scaling.

% \newpage

\balance
\bibliographystyle{ACM-Reference-Format}
\bibliography{sample-base-extend.bib}

%%% -*-BibTeX-*-
%%% Do NOT edit. File created by BibTeX with style
%%% ACM-Reference-Format-Journals [18-Jan-2012].

\begin{thebibliography}{52}

%%% ====================================================================
%%% NOTE TO THE USER: you can override these defaults by providing
%%% customized versions of any of these macros before the \bibliography
%%% command.  Each of them MUST provide its own final punctuation,
%%% except for \shownote{} and \showURL{}.  The latter two
%%% do not use final punctuation, in order to avoid confusing it with
%%% the Web address.
%%%
%%% To suppress output of a particular field, define its macro to expand
%%% to an empty string, or better, \unskip, like this:
%%%
%%% \newcommand{\showURL}[1]{\unskip}   % LaTeX syntax
%%%
%%% \def \showURL #1{\unskip}           % plain TeX syntax
%%%
%%% ====================================================================

\ifx \showCODEN    \undefined \def \showCODEN     #1{\unskip}     \fi
\ifx \showISBNx    \undefined \def \showISBNx     #1{\unskip}     \fi
\ifx \showISBNxiii \undefined \def \showISBNxiii  #1{\unskip}     \fi
\ifx \showISSN     \undefined \def \showISSN      #1{\unskip}     \fi
\ifx \showLCCN     \undefined \def \showLCCN      #1{\unskip}     \fi
\ifx \shownote     \undefined \def \shownote      #1{#1}          \fi
\ifx \showarticletitle \undefined \def \showarticletitle #1{#1}   \fi
\ifx \showURL      \undefined \def \showURL       {\relax}        \fi
% The following commands are used for tagged output and should be
% invisible to TeX
\providecommand\bibfield[2]{#2}
\providecommand\bibinfo[2]{#2}
\providecommand\natexlab[1]{#1}
\providecommand\showeprint[2][]{arXiv:#2}

\bibitem[Achiam et~al\mbox{.}(2023)]%
        {achiam2023gpt}
\bibfield{author}{\bibinfo{person}{Josh Achiam}, \bibinfo{person}{Steven Adler}, \bibinfo{person}{Sandhini Agarwal}, \bibinfo{person}{Lama Ahmad}, \bibinfo{person}{Ilge Akkaya}, \bibinfo{person}{Florencia~Leoni Aleman}, \bibinfo{person}{Diogo Almeida}, \bibinfo{person}{Janko Altenschmidt}, \bibinfo{person}{Sam Altman}, \bibinfo{person}{Shyamal Anadkat}, {et~al\mbox{.}}} \bibinfo{year}{2023}\natexlab{}.
\newblock \showarticletitle{Gpt-4 technical report}.
\newblock \bibinfo{journal}{\emph{arXiv}} (\bibinfo{year}{2023}).
\newblock


\bibitem[Andoni et~al\mbox{.}(2018)]%
        {andoni2018approximate}
\bibfield{author}{\bibinfo{person}{Alexandr Andoni}, \bibinfo{person}{Piotr Indyk}, {and} \bibinfo{person}{Ilya Razenshteyn}.} \bibinfo{year}{2018}\natexlab{}.
\newblock \showarticletitle{Approximate nearest neighbor search in high dimensions}. In \bibinfo{booktitle}{\emph{Proceedings of the International Congress of Mathematicians: Rio de Janeiro 2018}}. World Scientific, \bibinfo{pages}{3287--3318}.
\newblock


\bibitem[Cao et~al\mbox{.}(2025)]%
        {cao2025pantheon}
\bibfield{author}{\bibinfo{person}{Jiangxia Cao}, \bibinfo{person}{Pengbo Xu}, \bibinfo{person}{Yin Cheng}, \bibinfo{person}{Kaiwei Guo}, \bibinfo{person}{Jian Tang}, \bibinfo{person}{Shijun Wang}, \bibinfo{person}{Dewei Leng}, \bibinfo{person}{Shuang Yang}, \bibinfo{person}{Zhaojie Liu}, \bibinfo{person}{Yanan Niu}, {et~al\mbox{.}}} \bibinfo{year}{2025}\natexlab{}.
\newblock \showarticletitle{Pantheon: Personalized multi-objective ensemble sort via iterative pareto policy optimization}. In \bibinfo{booktitle}{\emph{Proceedings of the 34th ACM International Conference on Information and Knowledge Management}}. \bibinfo{pages}{5575--5582}.
\newblock


\bibitem[Cen et~al\mbox{.}(2020)]%
        {cen2020controllable}
\bibfield{author}{\bibinfo{person}{Yukuo Cen}, \bibinfo{person}{Jianwei Zhang}, \bibinfo{person}{Xu Zou}, \bibinfo{person}{Chang Zhou}, \bibinfo{person}{Hongxia Yang}, {and} \bibinfo{person}{Jie Tang}.} \bibinfo{year}{2020}\natexlab{}.
\newblock \showarticletitle{Controllable multi-interest framework for recommendation}. In \bibinfo{booktitle}{\emph{Proceedings of the 26th ACM SIGKDD international conference on knowledge discovery \& data mining}}. \bibinfo{pages}{2942--2951}.
\newblock


\bibitem[Chai et~al\mbox{.}(2025)]%
        {chai2025longer}
\bibfield{author}{\bibinfo{person}{Zheng Chai}, \bibinfo{person}{Qin Ren}, \bibinfo{person}{Xijun Xiao}, \bibinfo{person}{Huizhi Yang}, \bibinfo{person}{Bo Han}, \bibinfo{person}{Sijun Zhang}, \bibinfo{person}{Di Chen}, \bibinfo{person}{Hui Lu}, \bibinfo{person}{Wenlin Zhao}, \bibinfo{person}{Lele Yu}, {et~al\mbox{.}}} \bibinfo{year}{2025}\natexlab{}.
\newblock \showarticletitle{Longer: Scaling up long sequence modeling in industrial recommenders}. In \bibinfo{booktitle}{\emph{Proceedings of the Nineteenth ACM Conference on Recommender Systems}}. \bibinfo{pages}{247--256}.
\newblock


\bibitem[Chang et~al\mbox{.}(2023)]%
        {chang2023twin}
\bibfield{author}{\bibinfo{person}{Jianxin Chang}, \bibinfo{person}{Chenbin Zhang}, \bibinfo{person}{Zhiyi Fu}, \bibinfo{person}{Xiaoxue Zang}, \bibinfo{person}{Lin Guan}, \bibinfo{person}{Jing Lu}, \bibinfo{person}{Yiqun Hui}, \bibinfo{person}{Dewei Leng}, \bibinfo{person}{Yanan Niu}, \bibinfo{person}{Yang Song}, {et~al\mbox{.}}} \bibinfo{year}{2023}\natexlab{}.
\newblock \showarticletitle{TWIN: TWo-stage interest network for lifelong user behavior modeling in CTR prediction at kuaishou}. In \bibinfo{booktitle}{\emph{Proceedings of the 29th ACM SIGKDD Conference on Knowledge Discovery and Data Mining}}. \bibinfo{pages}{3785--3794}.
\newblock


\bibitem[Cheng et~al\mbox{.}(2025)]%
        {cheng2025choruscvr}
\bibfield{author}{\bibinfo{person}{Wei Cheng}, \bibinfo{person}{Yucheng Lu}, \bibinfo{person}{Boyang Xia}, \bibinfo{person}{Jiangxia Cao}, \bibinfo{person}{Kuan Xu}, \bibinfo{person}{Mingxing Wen}, \bibinfo{person}{Wei Jiang}, \bibinfo{person}{Jiaming Zhang}, \bibinfo{person}{Zhaojie Liu}, \bibinfo{person}{Liyin Hong}, {et~al\mbox{.}}} \bibinfo{year}{2025}\natexlab{}.
\newblock \showarticletitle{ChorusCVR: Chorus Supervision for Entire Space Post-Click Conversion Rate Modeling}.
\newblock \bibinfo{journal}{\emph{arXiv preprint arXiv:2502.08277}} (\bibinfo{year}{2025}).
\newblock


\bibitem[Covington et~al\mbox{.}(2016)]%
        {youtubednn}
\bibfield{author}{\bibinfo{person}{Paul Covington}, \bibinfo{person}{Jay Adams}, {and} \bibinfo{person}{Emre Sargin}.} \bibinfo{year}{2016}\natexlab{}.
\newblock \showarticletitle{Deep neural networks for youtube recommendations}. In \bibinfo{booktitle}{\emph{ACM Conference on Recommender Systems (RecSys)}}.
\newblock


\bibitem[Dao et~al\mbox{.}(2022)]%
        {dao2022flashattention}
\bibfield{author}{\bibinfo{person}{Tri Dao}, \bibinfo{person}{Dan Fu}, \bibinfo{person}{Stefano Ermon}, \bibinfo{person}{Atri Rudra}, {and} \bibinfo{person}{Christopher R{\'e}}.} \bibinfo{year}{2022}\natexlab{}.
\newblock \showarticletitle{Flashattention: Fast and memory-efficient exact attention with io-awareness}.
\newblock \bibinfo{journal}{\emph{Advances in neural information processing systems}}  \bibinfo{volume}{35} (\bibinfo{year}{2022}), \bibinfo{pages}{16344--16359}.
\newblock


\bibitem[Deng et~al\mbox{.}(2025)]%
        {deng2025onerec}
\bibfield{author}{\bibinfo{person}{Jiaxin Deng}, \bibinfo{person}{Shiyao Wang}, \bibinfo{person}{Kuo Cai}, \bibinfo{person}{Lejian Ren}, \bibinfo{person}{Qigen Hu}, \bibinfo{person}{Weifeng Ding}, \bibinfo{person}{Qiang Luo}, {and} \bibinfo{person}{Guorui Zhou}.} \bibinfo{year}{2025}\natexlab{}.
\newblock \showarticletitle{Onerec: Unifying retrieve and rank with generative recommender and iterative preference alignment}.
\newblock \bibinfo{journal}{\emph{arXiv preprint arXiv:2502.18965}} (\bibinfo{year}{2025}).
\newblock


\bibitem[Guan et~al\mbox{.}(2025)]%
        {guan2025make}
\bibfield{author}{\bibinfo{person}{Lin Guan}, \bibinfo{person}{Jia-Qi Yang}, \bibinfo{person}{Zhishan Zhao}, \bibinfo{person}{Beichuan Zhang}, \bibinfo{person}{Bo Sun}, \bibinfo{person}{Xuanyuan Luo}, \bibinfo{person}{Jinan Ni}, \bibinfo{person}{Xiaowen Li}, \bibinfo{person}{Yuhang Qi}, \bibinfo{person}{Zhifang Fan}, {et~al\mbox{.}}} \bibinfo{year}{2025}\natexlab{}.
\newblock \showarticletitle{Make It Long, Keep It Fast: End-to-End 10k-Sequence Modeling at Billion Scale on Douyin}.
\newblock \bibinfo{journal}{\emph{arXiv preprint arXiv:2511.06077}} (\bibinfo{year}{2025}).
\newblock


\bibitem[Guo et~al\mbox{.}(2025)]%
        {guo2025deepseek}
\bibfield{author}{\bibinfo{person}{Daya Guo}, \bibinfo{person}{Dejian Yang}, \bibinfo{person}{Haowei Zhang}, \bibinfo{person}{Junxiao Song}, \bibinfo{person}{Ruoyu Zhang}, \bibinfo{person}{Runxin Xu}, \bibinfo{person}{Qihao Zhu}, \bibinfo{person}{Shirong Ma}, \bibinfo{person}{Peiyi Wang}, \bibinfo{person}{Xiao Bi}, {et~al\mbox{.}}} \bibinfo{year}{2025}\natexlab{}.
\newblock \showarticletitle{Deepseek-r1: Incentivizing reasoning capability in llms via reinforcement learning}.
\newblock \bibinfo{journal}{\emph{arXiv preprint arXiv:2501.12948}} (\bibinfo{year}{2025}).
\newblock


\bibitem[Guo et~al\mbox{.}(2017)]%
        {guo2017deepfm}
\bibfield{author}{\bibinfo{person}{Huifeng Guo}, \bibinfo{person}{Ruiming Tang}, \bibinfo{person}{Yunming Ye}, \bibinfo{person}{Zhenguo Li}, {and} \bibinfo{person}{Xiuqiang He}.} \bibinfo{year}{2017}\natexlab{}.
\newblock \showarticletitle{DeepFM: a factorization-machine based neural network for CTR prediction}.
\newblock \bibinfo{journal}{\emph{arXiv preprint arXiv:1703.04247}} (\bibinfo{year}{2017}).
\newblock


\bibitem[Harlap et~al\mbox{.}(2018)]%
        {harlap2018pipedream}
\bibfield{author}{\bibinfo{person}{Aaron Harlap}, \bibinfo{person}{Deepak Narayanan}, \bibinfo{person}{Amar Phanishayee}, \bibinfo{person}{Vivek Seshadri}, \bibinfo{person}{Nikhil Devanur}, \bibinfo{person}{Greg Ganger}, {and} \bibinfo{person}{Phil Gibbons}.} \bibinfo{year}{2018}\natexlab{}.
\newblock \showarticletitle{Pipedream: Fast and efficient pipeline parallel dnn training}.
\newblock \bibinfo{journal}{\emph{arXiv preprint arXiv:1806.03377}} (\bibinfo{year}{2018}).
\newblock


\bibitem[Huang et~al\mbox{.}(2019)]%
        {huang2019gpipe}
\bibfield{author}{\bibinfo{person}{Yanping Huang}, \bibinfo{person}{Youlong Cheng}, \bibinfo{person}{Ankur Bapna}, \bibinfo{person}{Orhan Firat}, \bibinfo{person}{Dehao Chen}, \bibinfo{person}{Mia Chen}, \bibinfo{person}{HyoukJoong Lee}, \bibinfo{person}{Jiquan Ngiam}, \bibinfo{person}{Quoc~V Le}, \bibinfo{person}{Yonghui Wu}, {et~al\mbox{.}}} \bibinfo{year}{2019}\natexlab{}.
\newblock \showarticletitle{Gpipe: Efficient training of giant neural networks using pipeline parallelism}.
\newblock \bibinfo{journal}{\emph{Advances in neural information processing systems}}  \bibinfo{volume}{32} (\bibinfo{year}{2019}).
\newblock


\bibitem[Jacobs et~al\mbox{.}(2023)]%
        {jacobs2023deepspeed}
\bibfield{author}{\bibinfo{person}{Sam~Ade Jacobs}, \bibinfo{person}{Masahiro Tanaka}, \bibinfo{person}{Chengming Zhang}, \bibinfo{person}{Minjia Zhang}, \bibinfo{person}{Shuaiwen~Leon Song}, \bibinfo{person}{Samyam Rajbhandari}, {and} \bibinfo{person}{Yuxiong He}.} \bibinfo{year}{2023}\natexlab{}.
\newblock \showarticletitle{Deepspeed ulysses: System optimizations for enabling training of extreme long sequence transformer models}.
\newblock \bibinfo{journal}{\emph{arXiv preprint arXiv:2309.14509}} (\bibinfo{year}{2023}).
\newblock


\bibitem[Jiang et~al\mbox{.}(2024)]%
        {jiang2024megascale}
\bibfield{author}{\bibinfo{person}{Ziheng Jiang}, \bibinfo{person}{Haibin Lin}, \bibinfo{person}{Yinmin Zhong}, \bibinfo{person}{Qi Huang}, \bibinfo{person}{Yangrui Chen}, \bibinfo{person}{Zhi Zhang}, \bibinfo{person}{Yanghua Peng}, \bibinfo{person}{Xiang Li}, \bibinfo{person}{Cong Xie}, \bibinfo{person}{Shibiao Nong}, {et~al\mbox{.}}} \bibinfo{year}{2024}\natexlab{}.
\newblock \showarticletitle{$\{$MegaScale$\}$: Scaling large language model training to more than 10,000 $\{$GPUs$\}$}. In \bibinfo{booktitle}{\emph{21st USENIX Symposium on Networked Systems Design and Implementation (NSDI 24)}}. \bibinfo{pages}{745--760}.
\newblock


\bibitem[Kang and McAuley(2018)]%
        {sasrec}
\bibfield{author}{\bibinfo{person}{Wang-Cheng Kang} {and} \bibinfo{person}{Julian McAuley}.} \bibinfo{year}{2018}\natexlab{}.
\newblock \showarticletitle{Self-attentive sequential recommendation}. In \bibinfo{booktitle}{\emph{IEEE international conference on data mining (ICDM)}}.
\newblock


\bibitem[Korthikanti et~al\mbox{.}(2023)]%
        {korthikanti2023reducing}
\bibfield{author}{\bibinfo{person}{Vijay~Anand Korthikanti}, \bibinfo{person}{Jared Casper}, \bibinfo{person}{Sangkug Lym}, \bibinfo{person}{Lawrence McAfee}, \bibinfo{person}{Michael Andersch}, \bibinfo{person}{Mohammad Shoeybi}, {and} \bibinfo{person}{Bryan Catanzaro}.} \bibinfo{year}{2023}\natexlab{}.
\newblock \showarticletitle{Reducing activation recomputation in large transformer models}.
\newblock \bibinfo{journal}{\emph{Proceedings of Machine Learning and Systems}}  \bibinfo{volume}{5} (\bibinfo{year}{2023}), \bibinfo{pages}{341--353}.
\newblock


\bibitem[Kwon et~al\mbox{.}(2023)]%
        {kwon2023efficient}
\bibfield{author}{\bibinfo{person}{Woosuk Kwon}, \bibinfo{person}{Zhuohan Li}, \bibinfo{person}{Siyuan Zhuang}, \bibinfo{person}{Ying Sheng}, \bibinfo{person}{Lianmin Zheng}, \bibinfo{person}{Cody~Hao Yu}, \bibinfo{person}{Joseph~E. Gonzalez}, \bibinfo{person}{Hao Zhang}, {and} \bibinfo{person}{Ion Stoica}.} \bibinfo{year}{2023}\natexlab{}.
\newblock \showarticletitle{Efficient Memory Management for Large Language Model Serving with PagedAttention}. In \bibinfo{booktitle}{\emph{Proceedings of the ACM SIGOPS 29th Symposium on Operating Systems Principles}}.
\newblock


\bibitem[Li et~al\mbox{.}(2019)]%
        {li2019multi}
\bibfield{author}{\bibinfo{person}{Chao Li}, \bibinfo{person}{Zhiyuan Liu}, \bibinfo{person}{Mengmeng Wu}, \bibinfo{person}{Yuchi Xu}, \bibinfo{person}{Huan Zhao}, \bibinfo{person}{Pipei Huang}, \bibinfo{person}{Guoliang Kang}, \bibinfo{person}{Qiwei Chen}, \bibinfo{person}{Wei Li}, {and} \bibinfo{person}{Dik~Lun Lee}.} \bibinfo{year}{2019}\natexlab{}.
\newblock \showarticletitle{Multi-interest network with dynamic routing for recommendation at Tmall}. In \bibinfo{booktitle}{\emph{Proceedings of the 28th ACM international conference on information and knowledge management}}. \bibinfo{pages}{2615--2623}.
\newblock


\bibitem[Li et~al\mbox{.}(2023)]%
        {li2023blip}
\bibfield{author}{\bibinfo{person}{Junnan Li}, \bibinfo{person}{Dongxu Li}, \bibinfo{person}{Silvio Savarese}, {and} \bibinfo{person}{Steven Hoi}.} \bibinfo{year}{2023}\natexlab{}.
\newblock \showarticletitle{Blip-2: Bootstrapping language-image pre-training with frozen image encoders and large language models}. In \bibinfo{booktitle}{\emph{International conference on machine learning}}. PMLR, \bibinfo{pages}{19730--19742}.
\newblock


\bibitem[Li et~al\mbox{.}(2025)]%
        {li2025multi}
\bibfield{author}{\bibinfo{person}{Zihao Li}, \bibinfo{person}{Qiang Chen}, \bibinfo{person}{Lixin Zou}, \bibinfo{person}{Aixin Sun}, {and} \bibinfo{person}{Chenliang Li}.} \bibinfo{year}{2025}\natexlab{}.
\newblock \showarticletitle{Multi-Interest Recommendation: A Survey}.
\newblock \bibinfo{journal}{\emph{arXiv preprint arXiv:2506.15284}} (\bibinfo{year}{2025}).
\newblock


\bibitem[Liu et~al\mbox{.}(2024c)]%
        {liu2024deepseek}
\bibfield{author}{\bibinfo{person}{Aixin Liu}, \bibinfo{person}{Bei Feng}, \bibinfo{person}{Bing Xue}, \bibinfo{person}{Bingxuan Wang}, \bibinfo{person}{Bochao Wu}, \bibinfo{person}{Chengda Lu}, \bibinfo{person}{Chenggang Zhao}, \bibinfo{person}{Chengqi Deng}, \bibinfo{person}{Chenyu Zhang}, \bibinfo{person}{Chong Ruan}, {et~al\mbox{.}}} \bibinfo{year}{2024}\natexlab{c}.
\newblock \showarticletitle{Deepseek-v3 technical report}.
\newblock \bibinfo{journal}{\emph{arXiv preprint arXiv:2412.19437}} (\bibinfo{year}{2024}).
\newblock


\bibitem[Liu et~al\mbox{.}(2024a)]%
        {liu2024crm}
\bibfield{author}{\bibinfo{person}{Chi Liu}, \bibinfo{person}{Jiangxia Cao}, \bibinfo{person}{Rui Huang}, \bibinfo{person}{Kuo Cai}, \bibinfo{person}{Weifeng Ding}, \bibinfo{person}{Qiang Luo}, \bibinfo{person}{Kun Gai}, {and} \bibinfo{person}{Guorui Zhou}.} \bibinfo{year}{2024}\natexlab{a}.
\newblock \showarticletitle{CRM: Retrieval Model with Controllable Condition}.
\newblock \bibinfo{journal}{\emph{arXiv preprint arXiv:2412.13844}} (\bibinfo{year}{2024}).
\newblock


\bibitem[Liu et~al\mbox{.}(2024b)]%
        {kuaiformer}
\bibfield{author}{\bibinfo{person}{Chi Liu}, \bibinfo{person}{Jiangxia Cao}, \bibinfo{person}{Rui Huang}, \bibinfo{person}{Kai Zheng}, \bibinfo{person}{Qiang Luo}, \bibinfo{person}{Kun Gai}, {and} \bibinfo{person}{Guorui Zhou}.} \bibinfo{year}{2024}\natexlab{b}.
\newblock \showarticletitle{KuaiFormer: Transformer-Based Retrieval at Kuaishou}.
\newblock  (\bibinfo{year}{2024}).
\newblock


\bibitem[Luo et~al\mbox{.}(2024)]%
        {luo2024qarm}
\bibfield{author}{\bibinfo{person}{Xinchen Luo}, \bibinfo{person}{Jiangxia Cao}, \bibinfo{person}{Tianyu Sun}, \bibinfo{person}{Jinkai Yu}, \bibinfo{person}{Rui Huang}, \bibinfo{person}{Wei Yuan}, \bibinfo{person}{Hezheng Lin}, \bibinfo{person}{Yichen Zheng}, \bibinfo{person}{Shiyao Wang}, \bibinfo{person}{Qigen Hu}, {et~al\mbox{.}}} \bibinfo{year}{2024}\natexlab{}.
\newblock \showarticletitle{QARM: Quantitative Alignment Multi-Modal Recommendation at Kuaishou}.
\newblock \bibinfo{journal}{\emph{arXiv}} (\bibinfo{year}{2024}).
\newblock


\bibitem[Lv et~al\mbox{.}(2024)]%
        {lv2024marm}
\bibfield{author}{\bibinfo{person}{Xiao Lv}, \bibinfo{person}{Jiangxia Cao}, \bibinfo{person}{Shijie Guan}, \bibinfo{person}{Xiaoyou Zhou}, \bibinfo{person}{Zhiguang Qi}, \bibinfo{person}{Yaqiang Zang}, \bibinfo{person}{Ming Li}, \bibinfo{person}{Ben Wang}, \bibinfo{person}{Kun Gai}, {and} \bibinfo{person}{Guorui Zhou}.} \bibinfo{year}{2024}\natexlab{}.
\newblock \showarticletitle{MARM: Unlocking the Future of Recommendation Systems through Memory Augmentation and Scalable Complexity}.
\newblock \bibinfo{journal}{\emph{arXiv preprint arXiv:2411.09425}} (\bibinfo{year}{2024}).
\newblock


\bibitem[Ma et~al\mbox{.}(2018)]%
        {ma2018modeling}
\bibfield{author}{\bibinfo{person}{Jiaqi Ma}, \bibinfo{person}{Zhe Zhao}, \bibinfo{person}{Xinyang Yi}, \bibinfo{person}{Jilin Chen}, \bibinfo{person}{Lichan Hong}, {and} \bibinfo{person}{Ed~H Chi}.} \bibinfo{year}{2018}\natexlab{}.
\newblock \showarticletitle{Modeling task relationships in multi-task learning with multi-gate mixture-of-experts}. In \bibinfo{booktitle}{\emph{Proceedings of the 24th ACM SIGKDD international conference on knowledge discovery \& data mining}}. \bibinfo{pages}{1930--1939}.
\newblock


\bibitem[Mentzer et~al\mbox{.}(2023)]%
        {mentzer2023finite}
\bibfield{author}{\bibinfo{person}{Fabian Mentzer}, \bibinfo{person}{David Minnen}, \bibinfo{person}{Eirikur Agustsson}, {and} \bibinfo{person}{Michael Tschannen}.} \bibinfo{year}{2023}\natexlab{}.
\newblock \showarticletitle{Finite scalar quantization: Vq-vae made simple}.
\newblock \bibinfo{journal}{\emph{arXiv preprint arXiv:2309.15505}} (\bibinfo{year}{2023}).
\newblock


\bibitem[Pi et~al\mbox{.}(2020)]%
        {pi2020search}
\bibfield{author}{\bibinfo{person}{Qi Pi}, \bibinfo{person}{Guorui Zhou}, \bibinfo{person}{Yujing Zhang}, \bibinfo{person}{Zhe Wang}, \bibinfo{person}{Lejian Ren}, \bibinfo{person}{Ying Fan}, \bibinfo{person}{Xiaoqiang Zhu}, {and} \bibinfo{person}{Kun Gai}.} \bibinfo{year}{2020}\natexlab{}.
\newblock \showarticletitle{Search-based user interest modeling with lifelong sequential behavior data for click-through rate prediction}. In \bibinfo{booktitle}{\emph{Proceedings of the 29th ACM International Conference on Information \& Knowledge Management}}. \bibinfo{pages}{2685--2692}.
\newblock


\bibitem[Radford et~al\mbox{.}(2019)]%
        {radford2019language}
\bibfield{author}{\bibinfo{person}{Alec Radford}, \bibinfo{person}{Jeffrey Wu}, \bibinfo{person}{Rewon Child}, \bibinfo{person}{David Luan}, \bibinfo{person}{Dario Amodei}, \bibinfo{person}{Ilya Sutskever}, {et~al\mbox{.}}} \bibinfo{year}{2019}\natexlab{}.
\newblock \showarticletitle{Language models are unsupervised multitask learners}.
\newblock \bibinfo{journal}{\emph{OpenAI blog}} \bibinfo{volume}{1}, \bibinfo{number}{8} (\bibinfo{year}{2019}), \bibinfo{pages}{9}.
\newblock


\bibitem[Rafailov et~al\mbox{.}(2023)]%
        {rafailov2023direct}
\bibfield{author}{\bibinfo{person}{Rafael Rafailov}, \bibinfo{person}{Archit Sharma}, \bibinfo{person}{Eric Mitchell}, \bibinfo{person}{Christopher~D Manning}, \bibinfo{person}{Stefano Ermon}, {and} \bibinfo{person}{Chelsea Finn}.} \bibinfo{year}{2023}\natexlab{}.
\newblock \showarticletitle{Direct preference optimization: Your language model is secretly a reward model}.
\newblock \bibinfo{journal}{\emph{Advances in neural information processing systems}}  \bibinfo{volume}{36} (\bibinfo{year}{2023}), \bibinfo{pages}{53728--53741}.
\newblock


\bibitem[Rajbhandari et~al\mbox{.}(2020)]%
        {rajbhandari2020zero}
\bibfield{author}{\bibinfo{person}{Samyam Rajbhandari}, \bibinfo{person}{Jeff Rasley}, \bibinfo{person}{Olatunji Ruwase}, {and} \bibinfo{person}{Yuxiong He}.} \bibinfo{year}{2020}\natexlab{}.
\newblock \showarticletitle{Zero: Memory optimizations toward training trillion parameter models}. In \bibinfo{booktitle}{\emph{SC20: International Conference for High Performance Computing, Networking, Storage and Analysis}}. IEEE, \bibinfo{pages}{1--16}.
\newblock


\bibitem[Rajput et~al\mbox{.}(2023)]%
        {rajput2023recommender}
\bibfield{author}{\bibinfo{person}{Shashank Rajput}, \bibinfo{person}{Nikhil Mehta}, \bibinfo{person}{Anima Singh}, \bibinfo{person}{Raghunandan Hulikal~Keshavan}, \bibinfo{person}{Trung Vu}, \bibinfo{person}{Lukasz Heldt}, \bibinfo{person}{Lichan Hong}, \bibinfo{person}{Yi Tay}, \bibinfo{person}{Vinh Tran}, \bibinfo{person}{Jonah Samost}, {et~al\mbox{.}}} \bibinfo{year}{2023}\natexlab{}.
\newblock \showarticletitle{Recommender systems with generative retrieval}.
\newblock \bibinfo{journal}{\emph{Advances in Neural Information Processing Systems}}  \bibinfo{volume}{36} (\bibinfo{year}{2023}), \bibinfo{pages}{10299--10315}.
\newblock


\bibitem[Schulman et~al\mbox{.}(2017)]%
        {schulman2017proximal}
\bibfield{author}{\bibinfo{person}{John Schulman}, \bibinfo{person}{Filip Wolski}, \bibinfo{person}{Prafulla Dhariwal}, \bibinfo{person}{Alec Radford}, {and} \bibinfo{person}{Oleg Klimov}.} \bibinfo{year}{2017}\natexlab{}.
\newblock \showarticletitle{Proximal policy optimization algorithms}.
\newblock \bibinfo{journal}{\emph{arXiv preprint arXiv:1707.06347}} (\bibinfo{year}{2017}).
\newblock


\bibitem[Shi et~al\mbox{.}(2025)]%
        {shi2025llada}
\bibfield{author}{\bibinfo{person}{Teng Shi}, \bibinfo{person}{Chenglei Shen}, \bibinfo{person}{Weijie Yu}, \bibinfo{person}{Shen Nie}, \bibinfo{person}{Chongxuan Li}, \bibinfo{person}{Xiao Zhang}, \bibinfo{person}{Ming He}, \bibinfo{person}{Yan Han}, {and} \bibinfo{person}{Jun Xu}.} \bibinfo{year}{2025}\natexlab{}.
\newblock \showarticletitle{LLaDA-Rec: Discrete Diffusion for Parallel Semantic ID Generation in Generative Recommendation}.
\newblock \bibinfo{journal}{\emph{arXiv preprint arXiv:2511.06254}} (\bibinfo{year}{2025}).
\newblock


\bibitem[Shoeybi et~al\mbox{.}(2019)]%
        {shoeybi2019megatron}
\bibfield{author}{\bibinfo{person}{Mohammad Shoeybi}, \bibinfo{person}{Mostofa Patwary}, \bibinfo{person}{Raul Puri}, \bibinfo{person}{Patrick LeGresley}, \bibinfo{person}{Jared Casper}, {and} \bibinfo{person}{Bryan Catanzaro}.} \bibinfo{year}{2019}\natexlab{}.
\newblock \showarticletitle{Megatron-lm: Training multi-billion parameter language models using model parallelism}.
\newblock \bibinfo{journal}{\emph{arXiv preprint arXiv:1909.08053}} (\bibinfo{year}{2019}).
\newblock


\bibitem[Song et~al\mbox{.}(2021)]%
        {song2021fast}
\bibfield{author}{\bibinfo{person}{Xinying Song}, \bibinfo{person}{Alex Salcianu}, \bibinfo{person}{Yang Song}, \bibinfo{person}{Dave Dopson}, {and} \bibinfo{person}{Denny Zhou}.} \bibinfo{year}{2021}\natexlab{}.
\newblock \showarticletitle{Fast wordpiece tokenization}. In \bibinfo{booktitle}{\emph{Proceedings of the 2021 conference on empirical methods in natural language processing}}. \bibinfo{pages}{2089--2103}.
\newblock


\bibitem[Team et~al\mbox{.}(2025b)]%
        {5team2025glm45agenticreasoningcoding}
\bibfield{author}{\bibinfo{person}{GLM Team}, \bibinfo{person}{Aohan Zeng}, \bibinfo{person}{Xin Lv}, \bibinfo{person}{Qinkai Zheng}, {et~al\mbox{.}}} \bibinfo{year}{2025}\natexlab{b}.
\newblock \bibinfo{title}{GLM-4.5: Agentic, Reasoning, and Coding (ARC) Foundation Models}.
\newblock
\showeprint[arxiv]{2508.06471}~[cs.CL]
\urldef\tempurl%
\url{https://arxiv.org/abs/2508.06471}
\showURL{%
\tempurl}


\bibitem[Team et~al\mbox{.}(2025a)]%
        {team2025kimi}
\bibfield{author}{\bibinfo{person}{Kimi Team}, \bibinfo{person}{Yifan Bai}, \bibinfo{person}{Yiping Bao}, \bibinfo{person}{Guanduo Chen}, \bibinfo{person}{Jiahao Chen}, \bibinfo{person}{Ningxin Chen}, \bibinfo{person}{Ruijue Chen}, \bibinfo{person}{Yanru Chen}, \bibinfo{person}{Yuankun Chen}, \bibinfo{person}{Yutian Chen}, {et~al\mbox{.}}} \bibinfo{year}{2025}\natexlab{a}.
\newblock \showarticletitle{Kimi k2: Open agentic intelligence}.
\newblock \bibinfo{journal}{\emph{arXiv preprint arXiv:2507.20534}} (\bibinfo{year}{2025}).
\newblock


\bibitem[Wang et~al\mbox{.}(2024)]%
        {home}
\bibfield{author}{\bibinfo{person}{Xu Wang}, \bibinfo{person}{Jiangxia Cao}, \bibinfo{person}{Zhiyi Fu}, \bibinfo{person}{Kun Gai}, {and} \bibinfo{person}{Guorui Zhou}.} \bibinfo{year}{2024}\natexlab{}.
\newblock \showarticletitle{HoME: Hierarchy of Multi-Gate Experts for Multi-Task Learning at Kuaishou}.
\newblock  (\bibinfo{year}{2024}).
\newblock


\bibitem[Wu et~al\mbox{.}(2025)]%
        {wu2025qwen}
\bibfield{author}{\bibinfo{person}{Chenfei Wu}, \bibinfo{person}{Jiahao Li}, \bibinfo{person}{Jingren Zhou}, \bibinfo{person}{Junyang Lin}, \bibinfo{person}{Kaiyuan Gao}, \bibinfo{person}{Kun Yan}, \bibinfo{person}{Sheng-ming Yin}, \bibinfo{person}{Shuai Bai}, \bibinfo{person}{Xiao Xu}, \bibinfo{person}{Yilei Chen}, {et~al\mbox{.}}} \bibinfo{year}{2025}\natexlab{}.
\newblock \showarticletitle{Qwen-image technical report}.
\newblock \bibinfo{journal}{\emph{arXiv preprint arXiv:2508.02324}} (\bibinfo{year}{2025}).
\newblock


\bibitem[Yang et~al\mbox{.}(2025)]%
        {yang2025qwen3}
\bibfield{author}{\bibinfo{person}{An Yang}, \bibinfo{person}{Anfeng Li}, \bibinfo{person}{Baosong Yang}, \bibinfo{person}{Beichen Zhang}, \bibinfo{person}{Binyuan Hui}, \bibinfo{person}{Bo Zheng}, \bibinfo{person}{Bowen Yu}, \bibinfo{person}{Chang Gao}, \bibinfo{person}{Chengen Huang}, \bibinfo{person}{Chenxu Lv}, {et~al\mbox{.}}} \bibinfo{year}{2025}\natexlab{}.
\newblock \showarticletitle{Qwen3 technical report}.
\newblock \bibinfo{journal}{\emph{arXiv preprint arXiv:2505.09388}} (\bibinfo{year}{2025}).
\newblock


\bibitem[Yi et~al\mbox{.}(2019)]%
        {yi2019sampling}
\bibfield{author}{\bibinfo{person}{Xinyang Yi}, \bibinfo{person}{Ji Yang}, \bibinfo{person}{Lichan Hong}, \bibinfo{person}{Derek~Zhiyuan Cheng}, \bibinfo{person}{Lukasz Heldt}, \bibinfo{person}{Aditee Kumthekar}, \bibinfo{person}{Zhe Zhao}, \bibinfo{person}{Li Wei}, {and} \bibinfo{person}{Ed Chi}.} \bibinfo{year}{2019}\natexlab{}.
\newblock \showarticletitle{Sampling-bias-corrected neural modeling for large corpus item recommendations}. In \bibinfo{booktitle}{\emph{Proceedings of the 13th ACM conference on recommender systems}}. \bibinfo{pages}{269--277}.
\newblock


\bibitem[Zhang et~al\mbox{.}(2022)]%
        {zhang2022towards}
\bibfield{author}{\bibinfo{person}{Zhao-Yu Zhang}, \bibinfo{person}{Xiang-Rong Sheng}, \bibinfo{person}{Yujing Zhang}, \bibinfo{person}{Biye Jiang}, \bibinfo{person}{Shuguang Han}, \bibinfo{person}{Hongbo Deng}, {and} \bibinfo{person}{Bo Zheng}.} \bibinfo{year}{2022}\natexlab{}.
\newblock \showarticletitle{Towards understanding the overfitting phenomenon of deep click-through rate models}. In \bibinfo{booktitle}{\emph{Proceedings of the 31st ACM international conference on information \& knowledge management}}. \bibinfo{pages}{2671--2680}.
\newblock


\bibitem[Zhao et~al\mbox{.}(2023)]%
        {zhao2023pytorch}
\bibfield{author}{\bibinfo{person}{Yanli Zhao}, \bibinfo{person}{Andrew Gu}, \bibinfo{person}{Rohan Varma}, \bibinfo{person}{Liang Luo}, \bibinfo{person}{Chien-Chin Huang}, \bibinfo{person}{Min Xu}, \bibinfo{person}{Less Wright}, \bibinfo{person}{Hamid Shojanazeri}, \bibinfo{person}{Myle Ott}, \bibinfo{person}{Sam Shleifer}, {et~al\mbox{.}}} \bibinfo{year}{2023}\natexlab{}.
\newblock \showarticletitle{Pytorch fsdp: experiences on scaling fully sharded data parallel}.
\newblock \bibinfo{journal}{\emph{arXiv preprint arXiv:2304.11277}} (\bibinfo{year}{2023}).
\newblock


\bibitem[Zheng et~al\mbox{.}(2025)]%
        {zheng2025group}
\bibfield{author}{\bibinfo{person}{Chujie Zheng}, \bibinfo{person}{Shixuan Liu}, \bibinfo{person}{Mingze Li}, \bibinfo{person}{Xiong-Hui Chen}, \bibinfo{person}{Bowen Yu}, \bibinfo{person}{Chang Gao}, \bibinfo{person}{Kai Dang}, \bibinfo{person}{Yuqiong Liu}, \bibinfo{person}{Rui Men}, \bibinfo{person}{An Yang}, {et~al\mbox{.}}} \bibinfo{year}{2025}\natexlab{}.
\newblock \showarticletitle{Group sequence policy optimization}.
\newblock \bibinfo{journal}{\emph{arXiv preprint arXiv:2507.18071}} (\bibinfo{year}{2025}).
\newblock


\bibitem[Zheng et~al\mbox{.}(2024)]%
        {zheng2024sglang}
\bibfield{author}{\bibinfo{person}{Lianmin Zheng}, \bibinfo{person}{Liangsheng Yin}, \bibinfo{person}{Zhiqiang Xie}, \bibinfo{person}{Chuyue~Livia Sun}, \bibinfo{person}{Jeff Huang}, \bibinfo{person}{Cody~Hao Yu}, \bibinfo{person}{Shiyi Cao}, \bibinfo{person}{Christos Kozyrakis}, \bibinfo{person}{Ion Stoica}, \bibinfo{person}{Joseph~E Gonzalez}, {et~al\mbox{.}}} \bibinfo{year}{2024}\natexlab{}.
\newblock \showarticletitle{Sglang: Efficient execution of structured language model programs}.
\newblock \bibinfo{journal}{\emph{Advances in neural information processing systems}}  \bibinfo{volume}{37} (\bibinfo{year}{2024}), \bibinfo{pages}{62557--62583}.
\newblock


\bibitem[Zhou et~al\mbox{.}(2025)]%
        {zhou2025onerec}
\bibfield{author}{\bibinfo{person}{Guorui Zhou}, \bibinfo{person}{Jiaxin Deng}, \bibinfo{person}{Jinghao Zhang}, \bibinfo{person}{Kuo Cai}, \bibinfo{person}{Lejian Ren}, \bibinfo{person}{Qiang Luo}, \bibinfo{person}{Qianqian Wang}, \bibinfo{person}{Qigen Hu}, \bibinfo{person}{Rui Huang}, \bibinfo{person}{Shiyao Wang}, {et~al\mbox{.}}} \bibinfo{year}{2025}\natexlab{}.
\newblock \showarticletitle{OneRec Technical Report}.
\newblock \bibinfo{journal}{\emph{arXiv preprint arXiv:2506.13695}} (\bibinfo{year}{2025}).
\newblock


\bibitem[Zhou et~al\mbox{.}(2018)]%
        {zhou2018deep}
\bibfield{author}{\bibinfo{person}{Guorui Zhou}, \bibinfo{person}{Xiaoqiang Zhu}, \bibinfo{person}{Chenru Song}, \bibinfo{person}{Ying Fan}, \bibinfo{person}{Han Zhu}, \bibinfo{person}{Xiao Ma}, \bibinfo{person}{Yanghui Yan}, \bibinfo{person}{Junqi Jin}, \bibinfo{person}{Han Li}, {and} \bibinfo{person}{Kun Gai}.} \bibinfo{year}{2018}\natexlab{}.
\newblock \showarticletitle{Deep interest network for click-through rate prediction}. In \bibinfo{booktitle}{\emph{Proceedings of the 24th ACM SIGKDD international conference on knowledge discovery \& data mining}}. \bibinfo{pages}{1059--1068}.
\newblock


\bibitem[Zhu et~al\mbox{.}(2025)]%
        {zhu2025rankmixer}
\bibfield{author}{\bibinfo{person}{Jie Zhu}, \bibinfo{person}{Zhifang Fan}, \bibinfo{person}{Xiaoxie Zhu}, \bibinfo{person}{Yuchen Jiang}, \bibinfo{person}{Hangyu Wang}, \bibinfo{person}{Xintian Han}, \bibinfo{person}{Haoran Ding}, \bibinfo{person}{Xinmin Wang}, \bibinfo{person}{Wenlin Zhao}, \bibinfo{person}{Zhen Gong}, {et~al\mbox{.}}} \bibinfo{year}{2025}\natexlab{}.
\newblock \showarticletitle{Rankmixer: Scaling up ranking models in industrial recommenders}. In \bibinfo{booktitle}{\emph{Proceedings of the 34th ACM International Conference on Information and Knowledge Management}}. \bibinfo{pages}{6309--6316}.
\newblock


\end{thebibliography}

% \section{Appendix}

\begin{figure*}[t!]
  \centering
  \includegraphics[width=16cm,height=21cm]{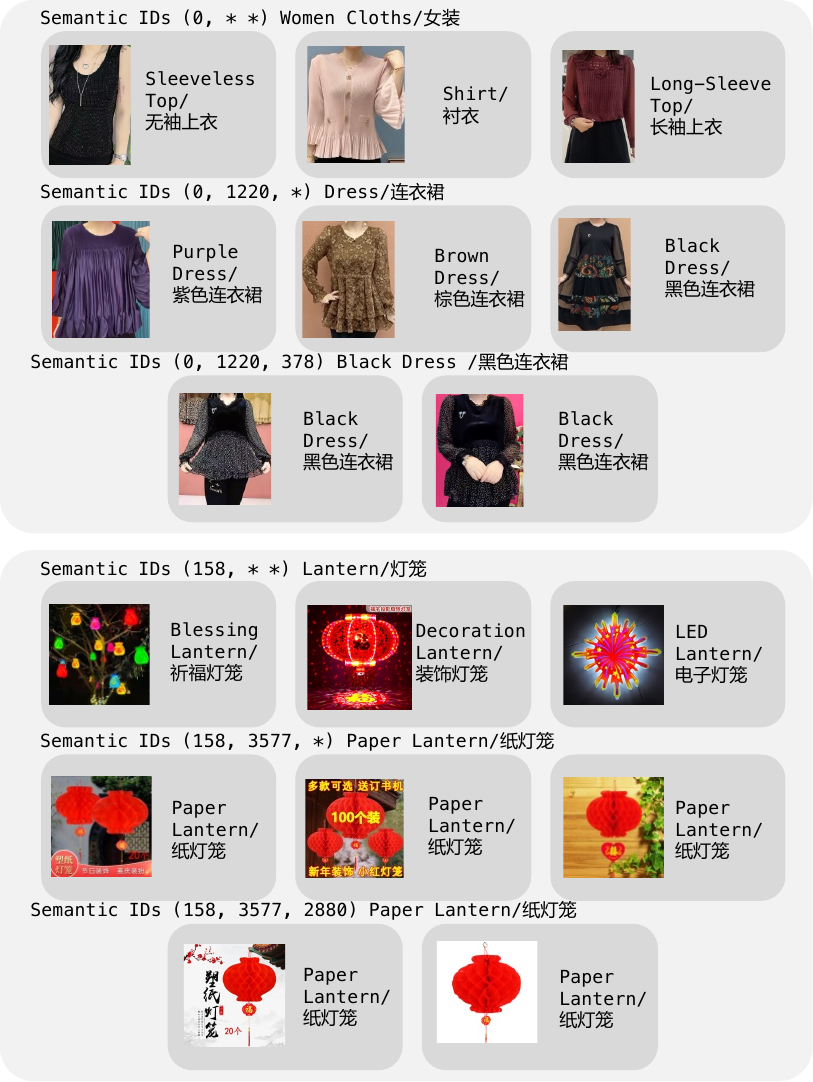}
  % \caption{Tokenizer Case Study at Product-Card.}
  \label{tokencase}
\end{figure*}

\end{document}